\documentclass[12pt]{article}
\usepackage[a4paper, top=16mm, text={200mm, 248mm}, includehead, includefoot, hmarginratio=1:1, heightrounded]{geometry}
\usepackage{amsmath,amssymb,mathrsfs,amsthm,tikz,shuffle,paralist}
\usepackage{dsfont}
\usepackage{color}
\definecolor{darkred}{RGB}{173,34,48}

\usetikzlibrary{snakes}
\usetikzlibrary{calc}
\usetikzlibrary{decorations}

\usepackage[all]{xypic}
\usepackage{jheppub}
\usepackage{hyperref}
\usepackage{subfigure}
\usepackage{caption}

\title{Schubert Problems, Positivity and Symbol Letters}

\date{\today}

 \author[a,b]{Qinglin Yang}%
\affiliation[a]{CAS Key Laboratory of Theoretical Physics, Institute of Theoretical Physics, Chinese Academy of Sciences, Beijing 100190, China}
\affiliation[b]{School of Physical Sciences, University of Chinese Academy of Sciences, No.19A Yuquan Road, Beijing 100049, China}

\emailAdd{yangqinglin@itp.ac.cn}

\abstract{We propose a geometrical approach to generate symbol letters of amplitudes/integrals in planar $\mathcal{N}=4$ Super Yang-Mills theory, known as {\it Schubert problems}. Beginning with one-loop integrals, we find that intersections of lines in momentum twistor space are always ordered on a given line, once the external kinematics $\mathbf{Z}$ is in the positive region $G_+(4,n)$. Remarkably, cross-ratios of these ordered intersections on a line, which are guaranteed to be positive now, nicely coincide with symbol letters of corresponding Feynman integrals, whose positivity is then concluded directly from such geometrical configurations. In particular, we reproduce from this approach the $18$ multiplicative independent algebraic letters for $n=8$ amplitudes up to three loops. Finally, we generalize the discussion to  two-loop Schubert problems and, again from ordered points on a line, generate a new kind of algebraic letters which mix two distinct square roots together. They have been found recently in the alphabet of two-loop double-box integral with $n\geq9$, and they are expected to appear in amplitudes at $k+\ell\geq4$. }
\begin{document}

\maketitle
\section{Introduction}
Recent years tremendous progress has been made to understand the hidden mathematical structure of scattering amplitudes, especially for planar $\mathcal{N}=4$ Super Yang-Mills (SYM) theory. Among all these developments, the hexagon and the heptagon bootstrap program ({\it c.f.}~\cite{Dixon:2011pw,Dixon:2014xca,Dixon:2014iba,Drummond:2014ffa,Dixon:2015iva,Caron-Huot:2016owq,Dixon:2016nkn,Drummond:2018caf, Caron-Huot:2019vjl, Caron-Huot:2019bsq, Dixon:2020cnr} and a review~\cite{Caron-Huot:2020bkp}.) plays a notable role. The crucial assumption of the program is that, as dual conformal invariant (DCI) multiple polylogarithmic (MPL) functions of external data,  $n=6,7$ amplitudes have quite limited {\it symbol letters} in their {\it alphabet} \cite{Golden:2013xva} (9 for $n=6$ and $42$ for $n=7$). After the space of allowed MPL functions with correct alphabet has been determined, amplitudes can thus be fixed in the space following certain conditions like symmetries, physical limits and so on. Consequently, six-point amplitudes have been determined through $7$ and $6$ loops for MHV and NMHV cases, and seven-point amplitudes through $4$ loops for the two cases respectively. Mathematically, their symbol letters are related to {\it $\mathcal{A}$-coordinates} of $G(4,n)$ \cite{Arkani-Hamed:2016byb} {\it cluster algebras}  \cite{fomin2002cluster,fomin2003cluster,berenstein2005cluster,fomin2007cluster}, and the $9$ or $42$ letters are thus explained by $G(4,6)\sim A_3$ and $G(4,7)\sim E_6$ algebras. For higher multiplicities, cluster bootstrap was obstructed since  cluster algebras $G(4,n)$ are of infinite type when $n\geq8$.  Moreover, explicit data of $n\geq8$ amplitudes ($n=8$ $\ell=2$ NMHV \cite{Zhang:2019vnm} and $\ell=3$ MHV \cite{Li:2021bwg}, also the $n=9$ case \cite{He:2020vob}) computed from the $\bar Q$ equation \cite{CaronHuot:2011kk} show that they involve symbol letters that are not rational functions of external data, which are not $\mathcal{A}$-coordinates of $G(4,n)$ as well.  Explanations on their letters have also been made from various approaches, such as Landau singularities \cite{Dennen:2015bet,Dennen:2016mdk,Prlina:2017tvx}, tropical positive Grassmannian \cite{speyer2005tropical,Drummond:2019qjk, Drummond:2019cxm, Henke:2019hve, Arkani-Hamed:2019rds,Arkani-Hamed:2020cig, Herderschee:2021dez,Henke:2021avn,Ren:2021ztg} and Yangian invariants and their associated Yangian letters \cite{Mago:2020kmp, He:2020uhb,Mago:2020nuv,Mago:2021luw}.

On the other hand, many impressive ideas and powerful tools such as differential equations \cite{Drummond:2010cz,Henn:2013pwa,Henn:2014qga}, bootstrap strategies \cite{Henn:2018cdp}  and Wilson-loop ${\rm d}\log$ form \cite{He:2020uxy,He:2020lcu} have been developed for the study of DCI integrals (see also \cite{Arkani-Hamed:2010pyv,Caron-Huot:2018dsv,Bourjaily:2018aeq,Herrmann:2019upk}) in planar $\mathcal{N}=4$ SYM theory as well, which have proved to be an ideal laboratory for exploring general Feynman integrals in QFT. One of the important progress is the discovery of cluster structures of individual integrals and their symbol letters, not only for $n=6,7$  DCI integrals \cite{Caron-Huot:2018dsv, Drummond:2017ssj}, but also for the cases beyond $\mathcal{N}=4$ SYM theory \cite{Chicherin:2020umh}. Most recently, in a series of works on DCI integrals, we revealed the connection between DCI kinematics and cluster algebras. After generating alphabets of certain integrals from their corresponding cluster algebras such as ladder-type integrals \cite{Drummond:2010cz,He:2021esx}, the cluster bootstrap program was successfully applied to determine explicit results of individual integrals with algebraic letters \cite{He:2021non,He:2021eec}.  

In this note, we provide an alternative path to generate symbol letters of amplitudes/integrals geometrically from {\it Schubert problems}. In our new approach, after we determine certain intersecting lines in $\mathbb{P}^3$ following the leading singularities of integrals, symbol letters are interpreted as cross-ratios of their intersections. Moreover, the positivity of letters becomes a direct conclusion from the ordering of intersections on a line.  While the reason why these geometrical configurations are connected to physical singularities still remains unclear, this method turns out to be quite powerful, recovering both rational letters and algebraic letters of amplitudes and integrals.  Note that it is quite different from the tropicalization approach, where rational letters and algebraic letters have distinct generations. In our formalism, either rational letters or algebraic letters, and even mixed algebraic letters with two distinct square roots as we will introduce, are all constructed from similar geometrical configurations, which indicates the connection between various kinds of letters.

The paper is organized as follows. We will firstly review some basic notations and definitions, such as momentum twistors, symbol letters and Schubert problems, which we use throughout this note. In section $2$ we will begin with some one-loop examples. Firstly proposed by N. Arkani-Hamed in the conference \cite{conference}, intersections from these one-loop Schubert problems reproduce all one-loop symbol letters, whose positivity in the positive region $G_+(4,n)$ is associated to the ordering of intersections on a given line. Especially we will mention N. Arkani-Hamed's construction on external lines of the four-mass Schubert problem, and see that it gives us algebraic letters with definite sign. In section $3$, we will generalize this construction and investigate many configurations with ordered intersections on external lines, which generate the $18$  algebraic letters for $8$-point amplitudes up to three loops and prove their positivity. Finally in section $4$, we generalize the discussion to two-loop Schubert problems and introduce a new kind of algebraic letters, each of which contains two distinct square roots. They are symbol letters of the $9$-point double-box integral, and are believed to appear in planar $\mathcal{N}=4$ SYM amplitudes at higher $k+\ell$.

\subsection{Notations and review}
Recall that for $n$ ordered, on-shell momenta $p_i$ in planar amplitudes/integrals, it is convenient to introduce $n$ {\it momentum twistors }\cite{Hodges:2009hk} $\mathbf{Z}:=Z_i^A$, with $A=1\cdots4$, following the definition:
\[Z_i=(\lambda_i^\alpha,x_i^{\alpha{\dot\alpha}}\lambda_{i\alpha})\]
where dual coordinates $x_i$ are defined by $p_i=x_{i{+}1}-x_i$. Momentum twistors trivialize both the on-shell conditions $p_i^2=0$ and the momentum conservation, and the squared distance of two dual points reads
$(x_i{-}x_j)^2=\frac{\langle i{-}1ij{-}1j\rangle}{\langle i{-}1i\rangle\langle j{-}1j\rangle}$. Here Pl\"ucker $\langle ijkl\rangle$ is the basic $SL(4)$ invariant $\langle ijkl\rangle:=\epsilon_{ABCD}Z_i^AZ_j^BZ_k^CZ_l^D$. Each dual point $x_i$ is mapped to a line $(i{-}1i)$ in momentum twistor space, and loop momentum $\ell$ is related to a bitwistor $(AB)$ as well. Consequently, propagator $(\ell-x_i)^2$ is rewritten as $\frac{\langle ABi{-}1i\rangle}{\langle AB\rangle\langle i{-}1i\rangle}$. Finally, as a collection of $n$ momentum twistors, external kinematics lives in the top cell of {\it Grassmannian} $G(4,n)$ \cite{Arkani-Hamed:2016byb}. Throughout this note, we mainly focus on a specific region, known as the {\it positive region} $G_+(4,n)$ in the whole configuration space, which is defined by  $\langle ijkl\rangle>0$ for arbitrary $i<j<k<l$. These positive conditions guarantee the positivity of symbol letters, as we will see. 

Throughout this note, integrals/amplitudes we take into account are DCI MPL functions of external data. Recall that  the total differential of a weight $w$ MPL function yields a general form as
\[{\rm d}\mathcal{F}^{w}=\sum_i\mathcal{F}_i^{(w{-}1)}{\rm d}\log x_i\]
Its {\it symbol} \cite{Goncharov:2010jf, Duhr:2011zq} is correspondingly defined as 
\[\mathcal{S}(\mathcal{F}^{w})=\sum_i\mathcal{S}(\mathcal{F}_i^{(w{-}1)})\otimes x_i\]
iteratively. Symbol of a weight $w$ MPL function is a sum over tensors with length $w$, whose entries are called its {\it symbol letters} and they are the main interests of this paper.

Finally, we review basic definitions of one-loop Schubert problems. Before it, as our most important example in this paper, let's review basic facts about the one-loop four-mass scalar box integral $F(i,j,k,l)$ and its alphabet, whose integrand in both dual coordinates and momentum twistors reads
\begin{align}\label{integrand1}
    \begin{tikzpicture}[baseline={([yshift=-.5ex]current bounding box.center)},scale=0.15]
                \draw[black,thick] (0,5)--(-5,5)--(-5,0)--(0,0)--cycle;
                \draw[black,thick] (1.93,5.52)--(0,5)--(0.52,6.93);
                \draw[black,thick] (1.93,-0.52)--(0,0)--(0.52,-1.93);
                \draw[black,thick] (-6.93,5.52)--(-5,5)--(-5.52,6.93);
                \draw[black,thick] (-6.93,-0.52)--(-5,0)--(-5.52,-1.93);
                \filldraw[black] (1.93,6) node[anchor=west] {{$j{-}1$}};
                \filldraw[black] (0.52,6.93) node[anchor=south] {{$i$}};
                \filldraw[black] (1.93,-1) node[anchor=west] {{$j$}};
                \filldraw[black] (0.52,-1.93) node[anchor=north] {{$k{-}1$}};
                \filldraw[black] (-6.93,6) node[anchor=east] {{$l$}};
                \filldraw[black] (-5.52,6.93) node[anchor=south] {{$i{-}1$}};
                \filldraw[black] (-6.93,-1) node[anchor=east] {{$l{-}1$}};
                \filldraw[black] (-5.52,-1.93) node[anchor=north] {{$k$}};
            \end{tikzpicture}&:=\int{\rm d}^4\ell\frac{(x_i{-}x_k)^2(x_j{-}x_l)^2}{(\ell{-}x_i)^2(\ell{-}x_j)^2(\ell{-}x_k)^2(\ell{-}x_l)^2}\nonumber\\
           &=\int_{AB}\frac{\langle i{-}1ik{-}1k\rangle\langle j{-}1jl{-}1l\rangle}{\langle ABi{-}1i\rangle\langle  ABj{-}1j\rangle\langle  ABk{-}1k\rangle\langle  ABl{-}1l\rangle}
\end{align}
Suppose the four indices satisfy $i<j{-}1$ and so on. As a DCI integral,  it depends on $2$ independent cross-ratios in momentum twistors as
\[u{=}\frac{\langle i{-}1ij{-}1j\rangle \langle k{-}1kl{-}1l\rangle}{\langle i{-}1i k{-}1k\rangle\langle  j{-}1j l{-}1l\rangle},\ v{=}\frac{\langle i{-}1ik{-}1k\rangle \langle j{-}1jl{-}1l\rangle}{\langle i{-}1i k{-}1k\rangle\langle  j{-}1j l{-}1l\rangle}\]
and is well-known to be a weight-two MPL function, whose symbol reads
\begin{equation}\label{box}
   \frac1{2\Delta_{i,j,k,l}} \biggl(v\otimes \frac{z_{i,j,k,l}}{\bar z_{i,j,k,l}}+u\otimes \frac{1-\bar z_{i,j,k,l}}{1-z_{i,j,k,l}} \biggr)
\end{equation}
with the definition 
\begin{equation}\label{delta}
\Delta_{i,j,k,l}=\sqrt{(1{-}u{-}v)^2-4u v}
\end{equation} 
and $z_{i,j,k,l} \bar z_{i,j,k,l}=u,\ (1{-}z_{i,j,k,l})(1{-}\bar z_{i,j,k,l})=v$. 

We naturally encounter {\it Schubert problem} when we compute the leading singularity (LS) of this integral in momentum twistor space \cite{Arkani-Hamed:2010pyv,Bourjaily:2013mma}. 
After taking residues of the integrand at $\langle ABi{-}1i\rangle=\langle  ABj{-}1j\rangle=\langle  ABk{-}1k\rangle=\langle  ABl{-}1l\rangle=0$, we find the solution for loop momentum $(AB)$ and arrive at an algebraic function of external data, which is called the leading singularity of the integral. Note that square root ${\Delta_{i,j,k,l}}$ can be generated from its leading singularity as well, since we in fact have $LS\propto\frac1{\Delta_{i,j,k,l}}$ (see appendix A for more details). In momentum twistor space  (projectively in $\mathbb{P}^3$), each on-shell condition $\langle ABm{-}1m\rangle=0$  indicates that the line $(AB)$ intersects with $(m{-}1m)$. Therefore, locating loop momentum $(AB)$ 
is interpreted as a Schubert problem  geometrically, {\it i.e.} we look for all the lines that simultaneously intersect with four lines $(i{-}1i)$, $(j{-}1j)$, $(k{-}1k)$ and $(l{-}1l)$ in generic positions.
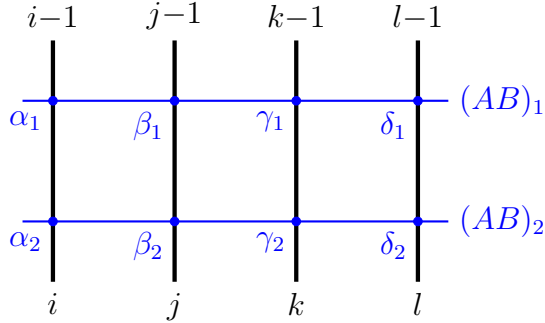
\begin{center}
\begin{tikzpicture}[scale=0.8]
\draw[black,ultra thick](-4,2)--(-4,-2);
\draw[black,ultra thick](-2,2)--(-2,-2);
\draw[black,ultra thick](0,2)--(0,-2);
\draw[black,ultra thick](2,2)--(2,-2);
\draw[blue,thick](-4.5,1)--(2.5,1);
\draw[blue,thick](-4.5,-1)--(2.5,-1);
\filldraw[blue] (2.5,1) node[anchor=west] {{$(AB)_1$}};
\filldraw[blue] (2.5,-1) node[anchor=west] {{$(AB)_2$}};
\filldraw[black] (-4,2) node[anchor=south] {{$i{-}1$}};
\filldraw[black] (-4,-2) node[anchor=north] {{$i$}};
\filldraw[black] (-2,2) node[anchor=south] {{$j{-}1$}};
\filldraw[black] (-2,-2) node[anchor=north] {{$j$}};
\filldraw[black] (0,2) node[anchor=south] {{$k{-}1$}};
\filldraw[black] (0,-2) node[anchor=north] {{$k$}};
\filldraw[black] (2,2) node[anchor=south] {{$l{-}1$}};
\filldraw[black] (2,-2) node[anchor=north] {{$l$}};
\filldraw[blue]  (-4,1) circle [radius=2pt];
\filldraw[blue]  (-4,-1) circle [radius=2pt];
\filldraw[blue]  (-2,1) circle [radius=2pt];
\filldraw[blue]  (-2,-1) circle [radius=2pt];
\filldraw[blue]  (0,1) circle [radius=2pt];
\filldraw[blue]  (0,-1) circle [radius=2pt];
\filldraw[blue]  (2,1) circle [radius=2pt];
\filldraw[blue]  (2,-1) circle [radius=2pt];
\filldraw[blue] (-4,1) node[anchor=north east] {{$\alpha_1$}};
\filldraw[blue] (-4,-1) node[anchor=north east] {{$\alpha_2$}};
\filldraw[blue] (-2,1) node[anchor=north east] {{$\beta_1$}};
\filldraw[blue] (-2,-1) node[anchor=north east] {{$\beta_2$}};
\filldraw[blue] (0,1) node[anchor=north east] {{$\gamma_1$}};
\filldraw[blue] (0,-1) node[anchor=north east] {{$\gamma_2$}};
\filldraw[blue] (2,1) node[anchor=north east] {{$\delta_1$}};
\filldraw[blue] (2,-1) node[anchor=north east] {{$\delta_2$}};
\end{tikzpicture}
\end{center}

Following the procedures in appendix A, we solve this Schubert problem and obtain  exactly two solutions, which are called $(AB)_1$ and $(AB)_2$ throughout this note. Intersection points $\{\alpha_i,\beta_i,\gamma_i,\delta_i\}$ on the two solutions can be fully parametried by external momentum twistors $Z_i$ as \eqref{four-mass-solutions} and \eqref{gammadelta} \cite{Bourjaily:2013mma}. Note that  $(1{-}u{-}v)^2{-}4 u v$ is positive definite once external data $\mathbf{Z}$ are evaluated in the positive region $G_+(4,n)$, {\it i.e.} $\langle ijkl\rangle>0$ for all $i<j<k<l$ \cite{Arkani-Hamed:2019rds}. Therefore the intersections \eqref{four-mass-solutions} and \eqref{gammadelta} involve only rational coefficients in the positive region, and $\{(AB)_i\}_{i=1,2}$ can be interpreted as lines in $\mathbb{P}^3$ geometrically.

A much more important observation is that, on each solution $(AB)_i$ with $i=1$ or $2$,  ordering of these four intersections $(\alpha_i,\beta_i,\gamma_i,\delta_i)$ is always fixed! Any two intersections are distinct and will never collide. Correspondingly, minors formed by any two points will have definite sign. In the following sections we will see that, this crucial property is satisfied not only in this four-mass Schubert problem case, but in various configurations throughout this note, and cross-ratios of intersections read symbol letters of corresponding amplitudes/integrals!

\section{Warm-up: one-loop Schubert problems and positivity}
In this warm-up section, we explore some one-loop configurations and their corresponding Schubert problems. In these one-loop examples, we will see that intersections from Schubert problems are always ordered on a given line. Furthermore, cross-ratios of these intersections reproduce DCI letters of one-loop amplitudes/integrals, and their positivity becomes a direct conclusion due to ordering of the intersections. Most examples in this section were firstly proposed by N.Arkani-Hamed in the conference \cite{conference}

\paragraph{Two-mass-easy box and the positivity of its letters}  Let's begin with a simple example, the two-mass-easy box $F(2,3,5,6)$ and its corresponding Schubert problem, 
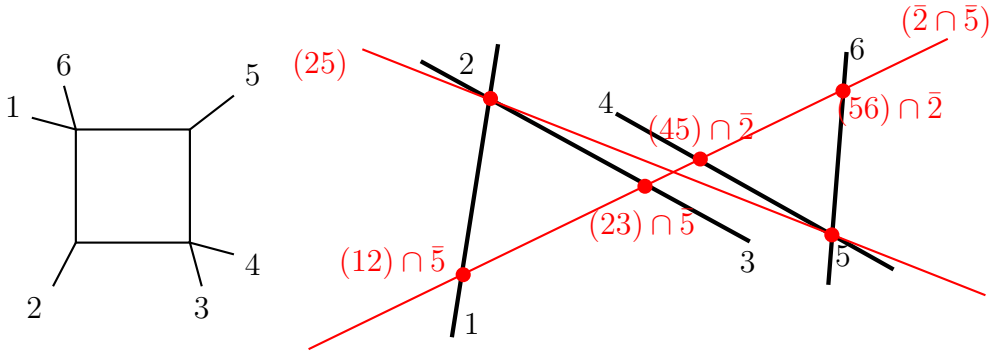
\begin{center}
\begin{tikzpicture}[baseline={([yshift=-12ex]current bounding box.center)},scale=0.3]
                \draw[black,thick] (0,5)--(-5,5)--(-5,0)--(0,0)--cycle;
                \draw[black,thick] (1.93,6.5)--(0,5);
                \draw[black,thick] (1.93,-0.52)--(0,0)--(0.52,-1.93);
                \draw[black,thick] (-6.93,5.52)--(-5,5)--(-5.52,6.93);
                \draw[black,thick] (-5,0)--(-6,-1.93);
                \filldraw[black] (1.93,6.5) node[anchor=south west] {{$5$}};
                \filldraw[black] (1.93,-1) node[anchor=west] {{$4$}};
                \filldraw[black] (0.52,-1.93) node[anchor=north] {{$3$}};
                \filldraw[black] (-6.93,6) node[anchor=east] {{$1$}};
                \filldraw[black] (-5.52,6.93) node[anchor=south] {{$6$}};
                \filldraw[black] (-6,-1.93) node[anchor=north east] {{$2$}};
            \end{tikzpicture}
\begin{tikzpicture}[scale=0.65]
\draw[black,ultra thick](-0.4788,4.1896)--(5.1391,1.0089);
\draw[black,ultra thick](3.8192,0.6953)--(4.188,5.4454);
\draw[black,ultra thick](-3.7979,-0.3736)--(-2.8629,5.6045);
\draw[black,ultra thick](-4.4282,5.2568)--(2.2296,1.5794);
\draw[red,thick](-5.6093,5.5009)--(6.9982,0.4829);
\draw[red,thick](-6.6967,-0.6177)--(6.2357,5.7135);
\node [fill=red,circle,inner sep=2pt] at (-3.5686,0.9021) {};
\node [fill=red,circle,inner sep=2pt] at (1.2283,3.2619) {};
\node [fill=red,circle,inner sep=2pt] at (4.117,4.6508) {};
\node [fill=red,circle,inner sep=2pt] at (0.109,2.7062) {};
\node [fill=red,circle,inner sep=2pt] at (-3.0172,4.4971) {};
\node [fill=red,circle,inner sep=2pt] at (3.8906,1.7142) {};
\node at (-3.5,5.2) {2};
\node at (-3.3955,-0.0915) {1};
\node at (2.1872,1.1686) {3};
\node at (-0.6773,4.4194) {4};
\node at (4.1172,1.2837) {5};
\node at (4.4083,5.5208) {6};
\node [red] at (-6.4662,5.1821) {$(25)$};
\node [red] at (6.1991,6.0584) {$(\bar2\cap\bar5)$};
\node [red] at (1.2332,3.8518) {$(45)\cap\bar2$};
\node [red] at (-5,1.1429) {$(12)\cap\bar5$};
\node [red] at (5.0556,4.2744) {$(56)\cap\bar2$};
\node [red] at (0.0396,1.9) {$(23)\cap\bar5$};
\end{tikzpicture}
\end{center}
Its leading singularity is supported by solutions of on-shell conditions $\langle AB12\rangle=\langle AB23\rangle=\langle AB45\rangle=\langle AB56\rangle=0$. Therefore we are looking for all possible $(AB)$ that intersect with four lines $\{(12),(23),(45),(56)\}$, which are $(AB)=(25)$ and $(AB)=(\bar2\cap\bar5)$, as the red lines in the figure. Here we denote $\bar i:=(i{-}1ii{+}1)$, and $(\bar i\cap\bar j)$ means the intersection line of two planes $(i{-}1ii{+}1)$ and $(j{-}1jj{+}1)$.  Let's explore the intersections on each line. Firstly, $(25)$ has only two intersections $2$ and $5$, which are of course distinct. Secondly, $(\bar2\cap\bar5)$ has four points 
\[\{(12)\cap\bar5,(23)\cap\bar5,(45)\cap\bar2,(56)\cap\bar2\}
\]
on it, where $(i{-}1i)\cap\bar j$ means the intersection point of line $(i{-}1i)$ with plane $(j{-}1jj{+}1)$.   To see they are ordered on $(\bar2\cap\bar5)$, we can set $A=(12)\cap\bar5$, $B=(23)\cap\bar5$ and  parametrize the rest two points  by $A$ and $B$ on the line. For example, projectively we have\footnote{Note that $(i{-}1i)\cap(j{-}1jj{+}1)=Z_{i{-}1}\langle ij{-}1j j{+}1\rangle-Z_{i}\langle i{-}1j{-}1j j{+}1\rangle=Z_{j{-}1}\langle i{-}1ijj{+}1\rangle-Z_{j}\langle i{-}1ij{-}1j{+}1\rangle+Z_{j{+}1}\langle i{-}1ij{-}1j\rangle$}
\begin{align*}
   (45)\cap(123) &\propto\langle2456\rangle(Z_1\langle2345\rangle-Z_2\langle1345\rangle+Z_3\langle1245\rangle) \\
   &=-\langle2345\rangle(Z_1\langle2456\rangle+Z_2\langle4561\rangle)+\langle1245\rangle(Z_3\langle3456\rangle+Z_2\langle4562\rangle)\\
   &=-\langle2345\rangle Z_A+\langle1245\rangle Z_B.
\end{align*}
Similar we have $(56)\cap(123)\propto-\langle2356\rangle Z_A+\langle1256\rangle Z_B$, and the four points form a $2\times4$ matrix as
\begin{equation}\label{2mematrix}
    \biggl(\begin{matrix}1&0&-\langle2345\rangle&-\langle2356\rangle\\0&1&\langle1245\rangle&\langle1256\rangle\end{matrix}\biggr)
\end{equation}
The upshot is that all of its $2\times2$ minors are positive definite when external $Z_i$s are in the positive region $G_+(4,6)$!  For instance, minor $(3,4)$ is $\langle1245\rangle\langle2356\rangle-\langle1256\rangle\langle2356\rangle=\langle1235\rangle\langle2456\rangle>0$, {\it etc}. We can also evaluate all the minors $(i,j)$ by cluster variables $\{f_i\}$ in \cite{He:2021eec}, and see that they are all positive polynomials. Therefore the four intersections are ordered on $(\bar2\cap\bar5)$.  Note that its $6$ minors $(i,j)$ for $1<i<j<6$ produce two non-trivial cross-ratios, which are DCI letters of the two-mass-easy box:
\begin{equation}\label{1}
\mathcal{U}:=\frac{(1,2)(3,4)}{(1,3)(2,4)}=1{-}U,\ \mathcal{V}:=\frac{(1,4)(2,3)}{(1,3)(2,4)}=U,\ \ U=\frac{\langle1256\rangle\langle2345\rangle}{\langle1245\rangle\langle2356\rangle}
\end{equation}
Ordering of the intersections guarantees that these cross-ratios are positive definite as well. 

Mathematically, the configuration space formed by $m$ ordered points on a line is called the {\it positive moduli space} $\mathcal{M}_{0,m}^+$, or $A_{m-3}$ configurations as they are also related to type-$A_n$ cluster algebras (see \cite{deligne1969irreducibility,devadoss1999tessellations,Arkani-Hamed:2017mur}). For general $m$, in total we can construct $\frac{m(m{-}3)}2$ cross-ratios from the line. Due to {\it u-equations} satisfied by these cross-ratios \cite{Arkani-Hamed:2017mur,Arkani-Hamed:2019plo,Arkani-Hamed:2020tuz}, only $m{-}3$ of them are independent. Our example corresponds to the case when $m=4$ especially, where we have $2$ cross-ratios $\{\mathcal{U},\mathcal{V}\}$ satisfying one relation $\mathcal{U}+\mathcal{V}=1$, and the positivity indicates $0<\mathcal{U}<1$ and $0<\mathcal{V}<1$. In fact, for general $A_{m-3}$, if the points are not allowed to collide, following from {\it u-equations} we always have $0<u<1$ for each cross-ratio $u$.


\paragraph{Four-mass box and the algebraic letters} 
Back to the four-mass case in review part, we can also choose two points on $(AB)_i$, parametrize the other points and write down the corresponding $2\times4$ matrix, which is much more complicated than \eqref{2mematrix} so we omit it here. Minors of arbitrary two points on $(AB)_i$ are algebraic functions of external data. Although implicit from the expressions directly, sign of the minors can be verified to be definite in the region $G_+(4,n)$. We reveal that on each $(AB)_i$, intersections are ordered as $\{\alpha_i,\beta_i,\gamma_i,\delta_i\}$, and the corresponding matrices are both positive definite. On the line $(AB)_1$, two non-trivial cross-ratios are
\[\mathcal{U}=z_{i,j,k,l},\ \  \mathcal{V}=1-z_{i,j,k,l}\]
Therefore we have $0<z_{i,j,k,l}<1$ in the positive region. Similarly four ordered points $(\alpha_2,\beta_2,\gamma_2,\delta_2)$ on $(AB)_2$  produce $\{\bar z_{i,j,k,l},1{-}\bar z_{i,j,k,l}\}$, which satisfy the condition $0<\bar z_{i,j,k,l}<1$. We see that from this configuration we successfully reproduce all the symbol letters in \eqref{box}! When one or more massive corners turn to be massless, four letters $\{z_{i,j,k,l},1-z_{i,j,k,l},\bar z_{i,j,k,l},1-\bar z_{i,j,k,l}\}$ degenerate to $\{U,1-U\}$ with certain cross-ratio $U$, which are letters of the corresponding lower-mass box. Hence from the one-loop Schubert problems, we can actually reproduce all one-loop letters in amplitudes/integrals and prove their positivity.

Finally, N. Arkani-Hamed has also suggested to consider the points on external lines $(i{-}1i)$ {\it etc}., and found out they also give certain algebraic letters for $n\geq8$ amplitudes \cite{Zhang:2019vnm,He:2020vob,Li:2021bwg}. Let's review this construction here as well. 
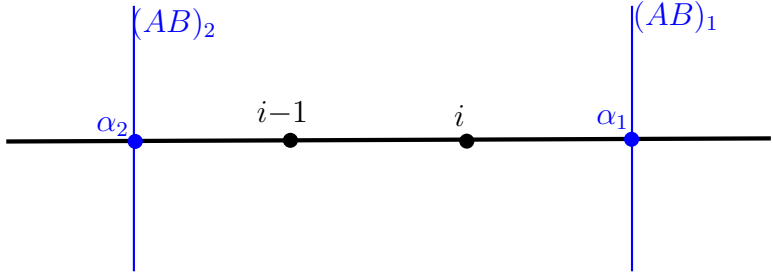
\begin{center}
    
\begin{tikzpicture}[scale=5]
\draw [black, ultra thick](4.8473,-6.6879) -- (6.8584,-6.6797);
\draw [blue,thick] (5.183,-6.3282) -- (5.183,-7.0322);
\draw [blue,thick] (6.4932,-6.3282) -- (6.4932,-7.0322);
\node [fill=blue,circle,inner sep=2pt] at (5.1864,-6.6878) {};
\node [fill=blue,circle,inner sep=2pt] at (6.4921,-6.682) {};
\node  [fill=black,circle,inner sep=2pt] at (5.5947,-6.6848) {};
\node  [fill=black,circle,inner sep=2pt] at (6.0586,-6.687) {};
\node [blue] at (5.1274,-6.6484) {$\alpha_2$};
\node [blue] at (6.4424,-6.6303) {$\alpha_1$};
\node at (5.574,-6.6131) {$i{-}1$};
\node at (6.0397,-6.6179) {$i$};
\node [blue] at (5.2876,-6.3768) {$(AB)_2$};
\node [blue] at (6.6085,-6.3581) {$(AB)_1$};
\end{tikzpicture}

\end{center}
As above, we notice that there are also four points $\{i{-}1,i,\alpha_1,\alpha_2\}$ on the external line $(i{-}1i)$. The upshot is that these four points are ordered as 
\[\{\alpha_2,i{-}1,i,\alpha_1\},\] 
which can be checked from the following positive definite matrix (parametrizing the four points by $i{-}1$ and $i$)
\begin{equation}
\biggl(\begin{matrix}
1&1&0&\partial_{Z_{i{-}1}}\alpha_1\\ \partial_{Z_i}\alpha_2&0&1&1
\end{matrix}\biggr)
\end{equation}
Here the notation $\partial_{Z_{i{-}1}}\alpha_1$ stands for the coefficient of $Z_{i{-}1}$ in the expression \eqref{A1} of $\alpha_1$. Especially, positive minor $(1,4)$ is proportional to the  square root $\Delta_{i,j,k,l}$. Furthermore, one of the cross-ratios from this $A_1$ configuration reads
\begin{equation}\label{oddletter1}
\tilde{L}_{i,j,k,l}=\frac{(1,2)(3,4)}{(1,3)(2,4)}=\frac{\chi_1-\bar z_{i,j,k,l}}{\chi_1- z_{i,j,k,l}},\ \chi_1=1+\frac{\langle (j{-}1j)\cap(i{-}1k{-}1k)i l{-}1l \rangle}{\langle i{-}1ik{-}1k\rangle\langle j{-}1jl{-}1l\rangle}
\end{equation}
It is a DCI algebraic function of external data  involving the square root $\Delta_{i,j,k,l}$. The crucial point is that it is contained in the alphabet of $n=8$ amplitudes up to three loops \cite{Zhang:2019vnm,Li:2021bwg} when $(i,j,k,l)=(2,4,6,8)$!  Similarly we can compute cross-ratios from lines $(j{-}1j)$ and so on, and conclude that they are all symbol letters with square root for amplitudes. 

These letters together with $\frac{z_{i,j,k,l}}{\bar z_{i,j,k,l}}$ and $\frac{1-z_{i,j,k,l}}{1-\bar z_{i,j,k,l}}$ are called {\it algebraic letters}, widely appearing in  amplitudes when $k+\ell\geq3$ and $n\geq8$, also in explicit results of individual Feynman integrals \cite{Bourjaily:2018aeq,He:2020lcu,He:2021non,He:2021eec}. For $n=8,9$ they are also recovered from topical fans of infinite-type cluster algebras $G(4,n)$ \cite{Drummond:2019qjk, Drummond:2019cxm, Henke:2019hve, Arkani-Hamed:2019rds,Arkani-Hamed:2020cig, Herderschee:2021dez,Henke:2021avn}, or from the ``four-mass-box" Yangian invariant and its associated Yangian letters \cite{Mago:2020kmp, He:2020uhb,Mago:2020nuv,Mago:2021luw}. Now \eqref{oddletter1} together with its new generation inspires us to reproduce all algebraic letters for $8$-point amplitudes by Schubert problems. In the following section we will generalize the idea of Schubert problems on external lines, and find out the $18$ algebraic letters for $8$-point amplitudes. We will also associate their positivity to the ordering of intersections.


\section{Schubert problems on external lines}
In this section we generalize the idea in the last section and see more configurations from Schubert problems, where intersections on given lines are always ordered, guaranteeing the positivity of minors formed by intersections. More explicitly, after four solutions from two different one-loop Schubert problems are determined, we consider the Schubert problem formed by these four lines. Several rational letters or algebraic letters of amplitudes/integrals can be reproduced from these configurations. Especially, such constructions are quite powerful, which enable us to reproduce the $18$ independent algebraic letters of $8$-point amplitudes up to three loops.

Let's review the algebraic letters of $8$-point amplitudes at first \cite{Zhang:2019vnm,Li:2021bwg}.  For $n=8$, there are two different four-mass topologies $F(1,3,5,7)$ and $F(2,4,6,8)$, respectively two different four-mass square roots $\Delta_{1,3,5,7}$ and $\Delta_{2,4,6,8}$. For each square root there are $9$ multiplicatively independent algebraic letters $L_i$ as (take $F(2,4,6,8)$ as an example)
\[L_i=\frac{\mathcal{X}_i-z_{2,4,6,8}}{\mathcal{X}_i-\bar z_{2,4,6,8}}\]
where
\[\biggl\{\mathcal{X}_1=0,\mathcal{X}_2=1,\mathcal{X}_3=\frac{\langle1236\rangle\langle8567\rangle}{\langle1256\rangle\langle8367\rangle},\mathcal{X}_7=\frac{\langle1246\rangle\langle8567\rangle}{\langle1256\rangle\langle8467\rangle}\biggr\},\]
together with their cyclic permutations ($L_3$ to $L_4, L_5,L_6$, and $L_7$ to $L_8,L_9,L_{10}$) under $Z_i\to Z_{i{+}2}$  (only $9$ of them are multiplicatively independent following the relation $\frac{L_5L_7L_9}{L_6L_8L_{10}}=1$). Note that algebraic letter \eqref{oddletter1} is related to $L_i$ as
\begin{equation}\label{oddletter12}
\tilde{L}_{2,4,6,8}=\frac{L_{10}}{L_4}
\end{equation}
Moreover, all $18$ independent algebraic letters are positive(negative) definite in the positive region $G_+(4,8)$, and can be recovered from Schubert problems, as will be  shown later.

\subsection{Combining two one-loop Schubert problems}
Let's go back to the final example in the last section and restrict our discussion at $(i,j,k,l)=(2,4,6,8)$.  Besides the two intersections $\{\alpha_1,\alpha_2\}$ produced by $(AB)_1$ and $(AB)_2$ on $(12)$, we also took $1$ and $2$ into account. 

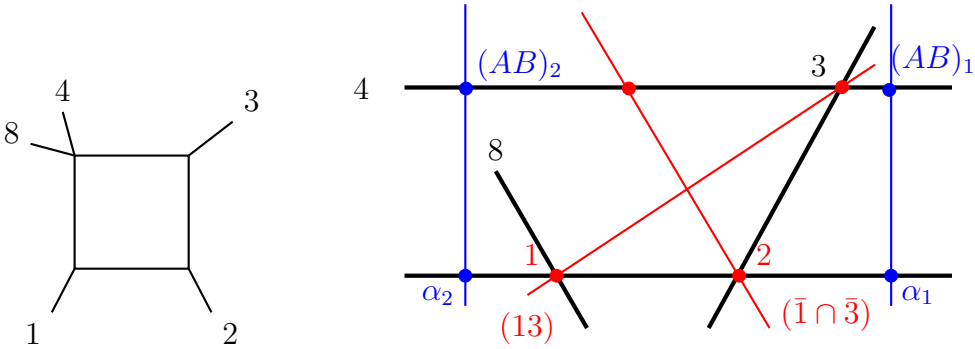
\begin{figure}[htbp]
\centering
     \begin{tikzpicture}[scale=0.3]
                \draw[black,thick] (0,5)--(-5,5)--(-5,0)--(0,0)--cycle;
                \draw[black,thick] (1.93,6.5)--(0,5);
                \draw[black,thick] (0,0)--(1,-1.93);
                \draw[black,thick] (-6.93,5.52)--(-5,5)--(-5.52,6.93);
                \draw[black,thick] (-5,0)--(-6,-1.93);
                \filldraw[black] (1.93,6.5) node[anchor=south west] {{$3$}};
                \filldraw[black] (1,-1.93) node[anchor=north west] {{$2$}};
                \filldraw[black] (-6.93,6) node[anchor=east] {{$8$}};
                \filldraw[black] (-5.52,6.93) node[anchor=south] {{$4$}};
                \filldraw[black] (-6,-1.93) node[anchor=north east] {{$1$}};
            \end{tikzpicture}
            \quad \quad
            \begin{tikzpicture}[scale=0.8]
\draw[black,ultra thick](-4.5,0)--(4.5,0);
\draw[black,ultra thick](-1.5,-0.867)--(-3,1.732);
\draw[black,ultra thick] (0.5,-0.867)--(3.2249,4.1255);
\draw[black,ultra thick] (4.5,3.12)--(-4.5,3.12);
\draw[red,thick] (3.2365,3.4994)--(-2.4786,-0.3195);
\draw[red,thick] (1.5,-0.866)--(-1.5875,4.3645);
\draw[blue,thick] (-3.5,-0.5)--(-3.5,4.5);
\draw[blue,thick] (3.5,-0.5)--(3.5,4.5);
\filldraw[red]  (-2,0) circle [radius=3pt];
\filldraw[red]  (2.6922,3.1263) circle [radius=3pt];
\filldraw[red]  (-0.8109,3.1028) circle [radius=3pt];
\filldraw[red]  (1,0) circle [radius=3pt];
\filldraw[blue]  (-3.5,0) circle [radius=3pt];
\filldraw[blue]  (3.5,0) circle [radius=3pt];
\filldraw[blue]  (3.4696,3.0793) circle [radius=3pt];
\filldraw[blue]  (-3.4896,3.1028) circle [radius=3pt];
\filldraw[blue] (-3.5,0) node[anchor=north east] {{$\alpha_2$}};
\filldraw[blue] (3.4883,0.0118) node[anchor=north west] {{$\alpha_1$}};
\filldraw[red] (-2.1,0) node[anchor=south east] {{$1$}};
\filldraw[red] (-2.4859,-0.4172) node[anchor=north] {{$(13)$}};
\filldraw[red] (1.5,-0.65) node[anchor=west] {{$(\bar1\cap\bar3)$}};
\filldraw[red] (1.1,0) node[anchor=south west] {{$2$}};
\filldraw[black] (2.6186,3.475) node[anchor=east] {{$3$}};
\filldraw[black] (-4.8972,3.0999) node[anchor=east] {{$4$}};
\filldraw[black] (-3,1.732) node[anchor=south] {{$8$}};
\filldraw[blue] (-3.5,3.5) node[anchor=west] {{$(AB)_2$}};
\filldraw[blue] (5.0752,3.5705) node[anchor=east] {{$(AB)_1$}};
\end{tikzpicture}
\caption{One-mass configuration (left) and four points on the line $(12)$ (right); the black lines $(81)$, $(12)$, $(23)$ and $(34)$ form a one-mass Schubert problem, whose solutions are the red lines, intersecting $(12)$ at $1$ and $2$; blue lines $(AB)_1$ and $(AB)_2$ are the solutions from the four-mass Schubert problem, intersecting with $(12)$ at $\alpha_1$ and $\alpha_2$.}\label{fig.one-mass}
\end{figure}

To generalize this construction, the crucial point is to interpret $1$ and $2$ as intersections on $(12)$ from a one-mass Schubert problem (Fig.\ref{fig.one-mass}).

Solutions of this problem are $(AB)_3=(13)$ and $(AB)_4=(\bar1\cap\bar3)$, intersecting $(12)$ at $1$ and $2$ respectively. Together with $(AB)_1$ and $(AB)_2$, four lines $\{(AB)_i\}_{i=1\cdots4}$ form a new Schubert problem, whose solutions are nothing but $(12)$ and $(34)$, since these two lines both intersect with all the four lines $\{(AB)_i\}_{i=1\cdots4}$. Finally, cross-ratio $\mathcal{U}=\frac{(1,2)(3,4)}{(1,3)(2,4)}$ of the four ordered intersections on $(12)$ reads the algebraic letter \eqref{oddletter12} we want.

This alternative viewpoint provides a natural way to generalize Schubert problems on external lines; we consider arbitrary two one-loop configurations sharing at least one external  $(i{-}1i)$. Four solutions $\{(AB)_i\}_{i=1\cdots4}$ from the two Schubert problems intersect $(i{-}1i)$ at four points respectively. Then a new Schubert problem formed by $\{(AB)_i\}_{i=1\cdots4}$ yields a solution $(i{-}1i)$. Four intersections on $(i{-}1i)$ will thus give us non-trivial, positive definite cross-ratios, once they are checked to be ordered.

As an illustration, consider the following two three-mass configurations at $n=8$ and their corresponding Schubert problems:
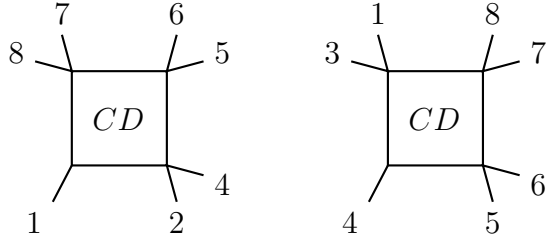
\begin{center}
    \begin{tikzpicture}[baseline={([yshift=-.5ex]current bounding box.center)},scale=0.25]
                \draw[black,thick] (0,5)--(-5,5)--(-5,0)--(0,0)--cycle;
                \draw[black,thick] (1.93,5.52)--(0,5)--(0.52,6.93);
                \draw[black,thick] (1.93,-0.52)--(0,0)--(0.52,-1.93);
                \draw[black,thick] (-6.93,5.52)--(-5,5)--(-5.52,6.93);
                \draw[black,thick] (-5,0)--(-6,-1.93);
                \filldraw[black] (1.93,6) node[anchor=west] {{$5$}};
                \filldraw[black] (0.52,6.93) node[anchor=south] {{$6$}};
                \filldraw[black] (1.93,-1) node[anchor=west] {{$4$}};
                \filldraw[black] (0.52,-1.93) node[anchor=north] {{$2$}};
                \filldraw[black] (-6.93,6) node[anchor=east] {{$8$}};
                \filldraw[black] (-5.52,6.93) node[anchor=south] {{$7$}};
                \filldraw[black] (-6,-1.93) node[anchor=north east] {{$1$}};
                \node at (-2.5477,2.4299) {$CD$};
            \end{tikzpicture}
            \quad\quad
                \begin{tikzpicture}[baseline={([yshift=-.5ex]current bounding box.center)},scale=0.25]
                \draw[black,thick] (0,5)--(-5,5)--(-5,0)--(0,0)--cycle;
                \draw[black,thick] (1.93,5.52)--(0,5)--(0.52,6.93);
                \draw[black,thick] (1.93,-0.52)--(0,0)--(0.52,-1.93);
                \draw[black,thick] (-6.93,5.52)--(-5,5)--(-5.52,6.93);
                \draw[black,thick] (-5,0)--(-6,-1.93);
                \filldraw[black] (1.93,6) node[anchor=west] {{$7$}};
                \filldraw[black] (0.52,6.93) node[anchor=south] {{$8$}};
                \filldraw[black] (1.93,-1) node[anchor=west] {{$6$}};
                \filldraw[black] (0.52,-1.93) node[anchor=north] {{$5$}};
                \filldraw[black] (-6.93,6) node[anchor=east] {{$3$}};
                \filldraw[black] (-5.52,6.93) node[anchor=south] {{$1$}};
                \filldraw[black] (-6,-1.93) node[anchor=north east] {{$4$}};
                \node at (-2.5477,2.4299) {$CD$};
            \end{tikzpicture}
\end{center}
There are four solutions for $(CD)$, which are $(CD)_1=((45)\cap\bar1,(67)\cap\bar1)$, $(CD)_2=(145)\cap(167)$ from the first configuration, and  $(CD)_3=((81)\cap\bar4,(67)\cap\bar4)$, $(CD)_4=(481)\cap(467)$ from the second. $(67)$ is one of the three lines that intersect with $\{(CD)_i\}_{i=1\cdots4}$ simultaneously \footnote{Note that here the four lines $\{(CD)_i\}_{i=1\cdots4}$ are not in generic positions, since they are all asked to intersect with 
$(81)$ $(45)$ and $(67)$ simultaneously by construction. Hence we obtain more than two solutions if we think of a Schubert problem formed by $\{(CD)_i\}_{i=1\cdots4}$. We will see a similar example in section $4$.}. Now we can check the four intersections from $\{(CD)_i\}_{i=1\cdots4}$ on $(67)$. For instance,  intersection of $(CD)_1$ with $(67)$ is just $(67)\cap\bar1$. As for $(CD)_2$, the intersection is $(67)\cap(145)$, since $(CD)_2$ itself is fully contained in the plane $(145)$, {\it etc.}. Finally, four intersections read $\{(67)\cap(145),(67)\cap(345),(67)\cap(812),(67)\cap(814)\}$, as shown in the figure below.
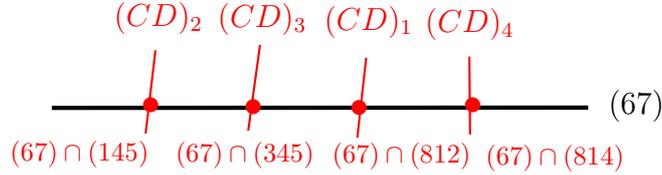
\begin{center}
 \begin{tikzpicture}[scale=1.5]
\draw [black, ultra thick](2.3852,-5.2281) -- (7.088,-5.2255);
\node [fill=red,circle,inner sep=2pt] at (6.0754,-5.2005) {};
\node [fill=red,circle,inner sep=2pt] at (5.0784,-5.2244) {};
\node [fill=red,circle,inner sep=2pt] at (4.1505,-5.2173) {};
\node [fill=red,circle,inner sep=2pt] at (3.2517,-5.2005) {};
\draw [red,thick](5.1415,-4.7856) -- (5.0636,-5.4922);
\draw [red,thick](3.294,-4.7186) -- (3.2077,-5.4554);
\draw [red,thick](4.2108,-4.672) -- (4.1075,-5.4297);
\draw [red,thick](6.0466,-4.7744) -- (6.052,-5.4635);
\node at (7.5098,-5.2129) {$(67)$};
\node [red] at (2.6267,-5.6395) {\footnotesize $(67)\cap(145)$};
\node [red] at (5.4501,-5.6568) {{\footnotesize $(67)\cap(812)$}};
\node [red] at (4.076,-5.6397) {\footnotesize$(67)\cap(345)$};
\node [red] at (6.7954,-5.6682) {\footnotesize $(67)\cap(814)$};
\node [red] at (3.3196,-4.4338) {$(CD)_2$};
\node [red] at (4.2296,-4.4412) {$(CD)_3$};
\node [red] at (5.1624,-4.464) {$(CD)_1$};
\node [red] at (6.0495,-4.4867) {$(CD)_4$};
\end{tikzpicture}
\end{center}
The upshot is that such four points are again ordered, as can be checked by evaluating minors of the following matrix (parametrizing four intersections by $6$ and $7$)
\begin{equation}
    \biggl(\begin{matrix}
    \langle7145\rangle& \langle7345\rangle&\langle7812\rangle&\langle7814\rangle\\\langle1456\rangle& \langle3456\rangle&\langle8126\rangle&\langle8146\rangle
    \end{matrix}\biggr)
\end{equation}
in the positive region $G_+(4,8)$. One of the corresponding cross-ratios reads:
\[
\mathcal{V}=\frac{\langle1458\rangle\langle1467\rangle\langle(\bar1\cap\bar4)67\rangle}{\langle1(28)(45)(67)\rangle\langle4(18)(35)(67)\rangle}
\]
Here we introduce the notation $\langle i(jk)(lm)(pq)\rangle=\langle ijlm\rangle\langle ikpq\rangle-\langle ijpq\rangle\langle iklm\rangle$. Notice that besides the ones in \eqref{box}, one-loop letters can only be $\{u,1{-}u\}$ with $u=\frac{x_{a,b}^2x_{c,d}^2}{x_{a,c}^2x_{b,d}^2}$, so factors $\langle1467\rangle$ and $\langle(\bar1\cap\bar4)67\rangle$ of $\mathcal{V}$ can never be factors of $8$-point one-loop letters. They are expected to appear in $\ell\geq2$, $n=8$ amplitudes/integrals. One of the simplest example is the penta-box ladder integrals \cite{Drummond:2010cz,Caron-Huot:2018dsv,He:2021esx}, whose Feynman diagrams read
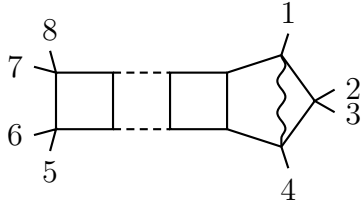
\begin{center}
\begin{tikzpicture}[baseline={([yshift=-.5ex]current bounding box.center)},scale=0.15]
                \draw[black,thick] (0,0)--(0,5)--(4.76,6.55)--(7.69,2.5)--(4.76,-1.55)--cycle;
                \draw[decorate, decoration=snake, segment length=12pt, segment amplitude=1.5pt, black,thick] (4.76,6.55)--(4.76,-1.55);
                \draw[black,thick] (9.43,1.5)--(7.69,2.5)--(9.43,3.5);
                \draw[black,thick] (4.76,6.55)--(5.37,8.45);
                \draw[black,thick] (4.76,-1.55)--(5.37,-3.45);
                \draw[black,thick] (0,5)--(-5,5)--(-5,0)--(0,0);
                \draw[thick,densely dashed] (-10,0) -- (-5,0);
                \draw[thick,densely dashed] (-10,5) -- (-5,5);
                \draw[black,thick] (-10,0)--(-10,5)--(-15,5)--(-15,0)--cycle;
                \draw[black,thick] (-16.93,5.52)--(-15,5)--(-15.52,6.93);
                \draw[black,thick] (-16.93,-0.52)--(-15,0)--(-15.52,-1.93);
                \filldraw[black] (5.37,8.45) node[anchor=south] {{$1$}};
                \filldraw[black] (5.37,-3.45) node[anchor=north] {{$4$}};
                \filldraw[black] (9.43,1.4) node[anchor=west] {{$3$}};
                \filldraw[black] (9.43,3.6) node[anchor=west] {{$2$}};
                \filldraw[black] (-16.93,5.52) node[anchor=east] {{$7$}};
                \filldraw[black] (-15.52,6.93) node[anchor=south] {{$8$}};
                \filldraw[black] (-16.93,-0.52) node[anchor=east] {{$6$}};
                \filldraw[black] (-15.52,-1.93) node[anchor=north] {{$5$}};
            \end{tikzpicture}
            \end{center}
They depend on three cross-ratios $u=(x_{17}^2 x_{25}^2)/(x_{15}^2 x_{27}^2)$,$ v=(x_{14}^2 x_{57}^2)/(x_{15}^2 x_{47}^2)$ and $w=(x_{15}^2 x_{24}^2)/(x_{14}^2 x_{25}^2)$. Up to all $\ell$, the penta-box ladder integrals have an alphabet as
\[\{u,v,w,1{-}u,1{-}v,1{-}w,1{-}u w,1{-}v w,1{-}u{-}v{+}u v w\}\]
following the explicit computation from Wilson-loop ${\rm d}\log$ form \cite{He:2020uxy}. The first $8$ symbol letters already appear in the alphabet of $\ell=1$ chiral pentagon integral, while the $9$-th letter only show up when $\ell\geq2$. In momentum twistor representation, the $9$-th letter reads
\begin{equation}\label{9letter}
1{-}u{-}v{+}u v w=\frac{\langle1467\rangle\langle(\bar1\cap\bar4)67\rangle}{\langle1267\rangle\langle1458\rangle\langle3467\rangle}
\end{equation}
consisting of the two factors we constructed from Schubert problem. In fact we have $\mathcal{V}=\frac{1-u-v+uvw}{(1-u)(1-v)}$.

\subsection{$8$-point algebraic letters from Schubert problems}
After discussions over the rational case, now we come back to $8$-point algebraic letters. The basic idea is the same; we are looking for two one-loop Schubert problems sharing an external line, whose four solutions produce four intersections on the line, which yield algebraic letters we want. It is easy to see that, to guarantee the resulting cross-ratios are algebraic, {\it i.e.} involving square root $\Delta_{2,4,6,8}$, we should fix one of the one-loop configurations as $F(2,4,6,8)$ and looking for the other one from possible lower-mass Schubert problems.

\begin{figure}[htbp]
\centering
            \begin{tikzpicture}[baseline={([yshift=-12ex]current bounding box.center)},scale=0.3]
                \draw[black,thick] (0,5)--(-5,5)--(-5,0)--(0,0)--cycle;
                \draw[black,thick] (1.93,6.5)--(0,5);
                \draw[black,thick] (1.93,-0.52)--(0,0)--(0.52,-1.93);
                \draw[black,thick] (-6.93,5.52)--(-5,5)--(-5.52,6.93);
                \draw[black,thick] (-5,0)--(-6,-1.93);
                \filldraw[black] (1.93,6.5) node[anchor=south west] {{$4$}};
                \filldraw[black] (1.93,-1) node[anchor=west] {{$3$}};
                \filldraw[black] (0.52,-1.93) node[anchor=north] {{$2$}};
                \filldraw[black] (-6.93,6) node[anchor=east] {{$8$}};
                \filldraw[black] (-5.52,6.93) node[anchor=south] {{$5$}};
                \filldraw[black] (-6,-1.93) node[anchor=north east] {{$1$}};
            \end{tikzpicture}
            \quad\quad
\begin{tikzpicture}[scale=0.8]
\draw [black, ultra thick](1.5,-6) -- (8.5,-6);
\draw [black, ultra thick] (2.9223,-4.9486) -- (6.7007,-7.6677);
\draw [black, ultra thick] (1.5,-3.5) -- (8.5,-3.5);
\draw[black, ultra thick]  (4.0862,-1.3844) -- (7.6932,-4.4075);
\draw [blue, thick](2,-6.5) -- (2,-3);
\draw [blue, thick](8,-6.5) -- (8,-3);
\draw [red,thick ](3.8641,-6.5698) -- (7.0898,-2.9774);
\draw [red,thick](5.2,-1) -- (5.2,-7.6);
\node [fill=red,circle,inner sep=2pt] at (4.4002,-6.0207) {};
\node [fill=red,circle,inner sep=2pt] at (5.2184,-2.3552) {};
\node [fill=red,circle,inner sep=2pt] at (5.2095,-6.6095) {};
\node [fill=red,circle,inner sep=2pt] at (5.2089,-3.5381) {};
\node [fill=red,circle,inner sep=2pt] at (6.629,-3.5004) {};
\node [fill=red,circle,inner sep=2pt] at (5.1901,-6.0302) {};
\node [fill=blue,circle,inner sep=2pt] at (7.9738,-6.0113) {};
\node [fill=blue,circle,inner sep=2pt] at (2.002,-6.0208) {};
\node [fill=blue,circle,inner sep=2pt] at (7.9927,-3.538) {};
\node [fill=blue,circle,inner sep=2pt] at (2.0209,-3.5191) {};
\node [red] at (4.3061,-5.5599) {1};
\node [red] at (6.5537,-3.0678) {4};
\node at (1.1463,-3.4252) {3};
\node at (7.0106,-7.8452) {8};
\node at (8.9008,-5.9737) {2};
\node at (3.9542,-1.0176) {5};
\node [blue] at (1.2583,-2.7188) {$(AB)_2$};
\node [blue] at (8.7355,-2.8905) {$(AB)_1$};
\node [red] at (6.1325,-5.5952) {$(12)\cap\bar4$};
\node [red] at (4.2296,-3.0128) {$(34)\cap\bar1$};
\node [blue] at (1.5467,-6.3768) {$\alpha_2$};
\node [blue] at (8.3742,-6.3486) {$\alpha_1$};
\node [blue]at (1.5749,-3.9035) {$\beta_2$};
\node [blue] at (8.346,-3.8283) {$\beta_1$};
\node [red] at (7.2569,-2.4957) {$(14)$};
\node [red] at (6.0232,-1.1832) {$(\bar1\cap\bar4)$};
\end{tikzpicture}
\caption{Two-mass-easy configuration (left) and four intersections on $(12)$ or $(34)$ (right); the black lines $(81)$, $(12)$, $(34)$ and $(45)$ form a two-mass-easy Schubert problem, whose solutions are the red lines, intersecting $(12)$ (as well as $(34)$) at two points. The blue lines are solutions from the four-mass Schubert problem again}            \label{fig.two-mass-easy} 
\end{figure}
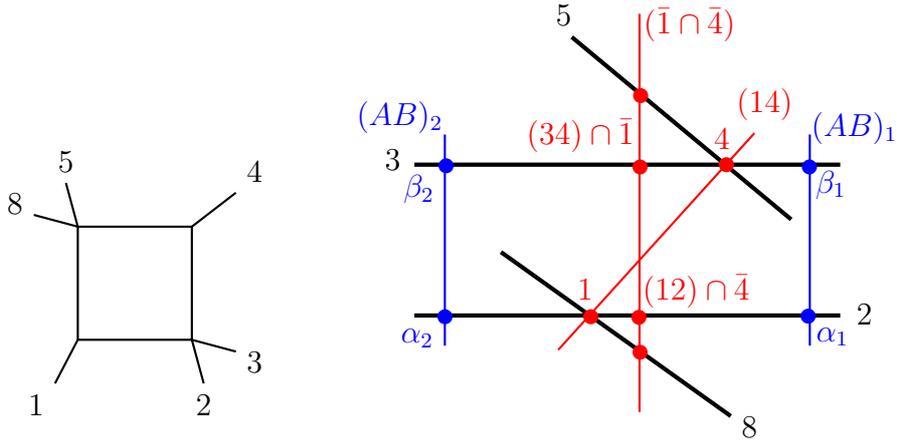

For instance, let's  consider a lower-mass Schubert problem as Fig.\ref{fig.two-mass-easy}.  Two lines $(AB)_1$ and $(AB)_2$ still come from the four-mass Schubert problem. For the lower-mass configuration, we look for $(AB)$ satisfying $\langle AB81\rangle=\langle AB12\rangle=\langle AB34\rangle=\langle AB45\rangle=0$, which are $(AB)_3=(14)$ and $(AB)_4=(\bar1\cap\bar4)$.
Now let's figure out intersections on $(12)$. (Note that $(12)$ and $(34)$ are solutions of the Schubert problem formed by $\{(AB)_i\}_{i=1\cdots4}$). We already have $\{\alpha_1,\alpha_2\}$ from $(AB)_1$ and $(AB)_2$. Moreover, $(AB)_3$ and $(AB)_4$ intersect with $(12)$ at $1$ and $(12)\cap(345)$ respectively. Such four points are ordered by checking  the following matrix
\begin{equation}\label{22}
\biggl(\begin{matrix}
1&1&{-}\langle2345\rangle&&\partial_{Z_1}\alpha_1\\ \partial_{Z_2}\alpha_2&0&{-}\langle3451\rangle&&1
\end{matrix}\biggr)
\end{equation}
is positive definite. Note that one of the non-trivial cross-ratios from this $A_1$  reads 
\begin{equation}\label{oddletter2}
   \mathcal{U}=\frac{(1,2)(3,4)}{(1,3)(2,4)}=\frac{L_2 L_5}{L_4L_6 L_8}
\end{equation}
giving an algebraic letter we want\footnote{Note that here we can also consider $\mathcal{V}$ from these $A_1$ configurations. However, it can be checked that $\mathcal{V}$ is a multiplicative combination of $\mathcal{U}$ and certain other factors like rational letters and $\Delta_{2,4,6,8}$, producing no new members in the $9$-dimensional multiplicative space.}. Note that $\mathcal{U}$ together with its three cyclic permutations by $Z_i\to Z_{i{+}2}$ are all members in the $9$-dimensional multiplicative space for $8$-point algebraic letters with $\Delta_{2,4,6,8}$. By exploring cyclic permutations of the configuration in Fig.\ref{fig.two-mass-easy}, we reproduce $3$ cyclic images of $\mathcal{U}$ correspondingly.

Other odd letters with $\Delta_{2,4,6,8}$ in the $9$-dimensional space can also be generated through similar approaches. Note that together with their cyclic permutations, letters $L_1$, $L_2$, \eqref{oddletter12}, and \eqref{oddletter2} provide $8$ independent members in the $9$-dimensional space. As for the final one, it can be generated from the following configuration 
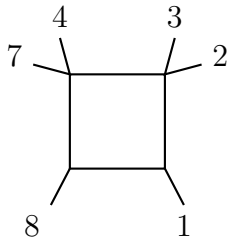
\begin{center}
     \begin{tikzpicture}[baseline={([yshift=-.5ex]current bounding box.center)},scale=0.25]
                \draw[black,thick] (0,5)--(-5,5)--(-5,0)--(0,0)--cycle;
                \draw[black,thick] (1.93,5.52)--(0,5)--(0.52,6.93);
                \draw[black,thick] (0,0)--(1,-1.93);
                \draw[black,thick] (-6.93,5.52)--(-5,5)--(-5.52,6.93);
                \draw[black,thick] (-5,0)--(-6,-1.93);
                \filldraw[black] (1.93,6) node[anchor=west] {{$2$}};
                \filldraw[black] (0.52,6.93) node[anchor=south] {{$3$}};
                \filldraw[black] (1,-1.93) node[anchor=north] {{$1$}};
                \filldraw[black] (-6.93,6) node[anchor=east] {{$7$}};
                \filldraw[black] (-5.52,6.93) node[anchor=south] {{$4$}};
                \filldraw[black] (-6,-1.93) node[anchor=north east] {{$8$}};
            \end{tikzpicture}
\end{center}
This two-mass-hard Schubert problem yields two solutions $(AB)_3=(1(34)\cap\bar8)$ and $(AB)_4=(8(34)\cap\bar1)$, intersecting $(12)$ at $(12)\cap(348)$ and $1$ respectively. Together with $\{\alpha_1,\alpha_2\}$ from the four-mass problem, four ordered points $\{(12)\cap(348),\alpha_1,1,\alpha_2\}$ on $(12)$ yield  an algebraic letter
\[\mathcal{U}/\mathcal{V}=\frac{L_5 L_9}{L_4L_6 L_8}\]
from its $A_1$ configuration. It contributes the rest letter we need to reproduce the full $9$-dimensional space. 

Remark that besides the configurations we went through, actually the $9$ letters can be obtained from certain similar constructions and corresponding $A_n$ configurations as well. Our approach is only one of the proper options. For instance, the $9$-dimensional space can also be recovered from ordered intersections on $(34)$ in all configurations we went through,  instead of $(12)$. It is just like the case in \cite{Henke:2019hve,Drummond:2019cxm}, where more than $9$ algebraic letters can be computed from limit rays of tropical $G(4,8)$ but only $9$ of them are multiplicatively independent.  Here we emphasis that different from the cluster algebra approach, where rational letters and algebraic letters are constructed from pretty different ways, from the viewpoint of Schubert problems we unify generations of these two kinds of letters, and their positivity becomes a direct conclusion from $A_1$ configurations. It is also an interesting question to generate all $272$ (or $356$) rational letters from tropical $G(4,8)$ through this procedure.

Let's leave one more comment on the positivity of algebraic letters. Note that unlike rational cases, it is sometimes a little intricate to analytically prove the positivity of algebraic letters $\frac{a+\Delta_{2,4,6,8}}{a-\Delta_{2,4,6,8}}$ (or $\frac{\Delta_{2,4,6,8}+a}{\Delta_{2,4,6,8}-a}$) directly from their Pl\"ucker representations.  Generally $a$ are not positive definite. An efficient way to achieve the goal is parametrizing  $a\pm\Delta_{2,4,6,8}$ by cluster variables $\{f_i\}$ and computing $a^2-\Delta_{2,4,6,8}^2$ \cite{He:2021eec}. The upshot is that throughout this note, expressions $a^2-\Delta_{2,4,6,8}^2$ are always positive (or negative) polynomials of $\{f_i\}$. Since in amplitudes/integrals, only combinations $\frac{a+\Delta_{2,4,6,8}}{a-\Delta_{2,4,6,8}}$ and  $a^2-\Delta_{2,4,6,8}^2$ appear as symbol letters, this in fact proves the positivity of corresponding algebraic letters directly.

\section{Mixed algebraic letters and the $9$-point double-box integral}
Finally, we look into some more non-trivial examples, which provide algebraic letters with more than one square root. We will see that such complicated algebraic letters can also be constructed from  almost the same configurations on external lines, which reveals the deep relation between different kinds of letters.

\paragraph{Mixed algebraic letters with two four-mass square roots} After considering four solutions from two one-loop Schubert problem with at most one four-mass configurations, there is nothing stopping us from combining two different four-mass Schubert problems and exploring intersections on external lines, which only realizes when $n\geq9$. 

\begin{figure}[htbp]
\centering
\begin{tikzpicture}[baseline={([yshift=-12ex]current bounding box.center)},scale=0.15]
                \draw[black,thick] (0,5)--(-5,5)--(-5,0)--(0,0)--cycle;
                \draw[black,thick] (1.93,5.52)--(0,5)--(0.52,6.93);
                \draw[black,thick] (1.93,-0.52)--(0,0)--(0.52,-1.93);
                \draw[black,thick] (-6.93,5.52)--(-5,5)--(-5.52,6.93);
                \draw[black,thick] (-6.93,-0.52)--(-5,0)--(-5.52,-1.93);
                \filldraw[black] (1.93,6) node[anchor=west] {{$4$}};
                \filldraw[black] (0.52,6.93) node[anchor=south] {{$3$}};
                \filldraw[black] (1.93,-1) node[anchor=west] {{$5$}};
                \filldraw[black] (0.52,-1.93) node[anchor=north] {{$6$}};
                \filldraw[black] (-6.93,6) node[anchor=east] {{$9$}};
                \filldraw[black] (-5.52,6.93) node[anchor=south] {{$2$}};
                \filldraw[black] (-6.93,-1) node[anchor=east] {{$8$}};
                \filldraw[black] (-5.52,-1.93) node[anchor=north] {{$7$}};
                \node at (-2.1034,-8) {$F(3,5,7,9)$};
            \end{tikzpicture}
            \quad\quad
\begin{tikzpicture}[scale=0.85]
\draw[blue,thick](-4,4.5)--(-4,0.5);
\draw[blue,thick](-0.0218,4.5155)--(-2,0.5);
\draw[blue,thick](-2.2699,4.5346)--(0,0.5);
\draw[blue,thick](2,4.5)--(2,0.5);
\draw[black,ultra thick](-4.5,1)--(2.5,1);
\draw[black,ultra thick](-4.5,2)--(2.5,2);
\draw[black,ultra thick](-4.5,3)--(2.5,3);
\draw[black,ultra thick](-4.4652,3.4834)--(-0.5494,4.6);
\draw[black,ultra thick](2.4401,3.4437)--(-1.5354,4.5954);
\filldraw[black] (2.5,1) node[anchor=west] {{$(67)$}};
\filldraw[blue] (-3.9967,4.5047) node[anchor=south] {{$(CD)_3$}};
\filldraw[blue] (-0.0155,4.525) node[anchor=south] {{$(CD)_1$}};
\filldraw[blue] (-2.2455,4.5435) node[anchor=south] {{$(CD)_4$}};
\filldraw[blue] (2.0224,4.5055) node[anchor=south] {{$(CD)_2$}};
\filldraw[blue]  (-4,1) circle [radius=2pt];
\filldraw[blue]  (-1.7454,1.0107) circle [radius=2pt];
\filldraw[blue]  (-0.2755,1.0034) circle [radius=2pt];
\filldraw[blue]  (2,1) circle [radius=2pt];
\filldraw[blue] (-4,1) node[anchor=north east] {{$\gamma_{21}$}};
\filldraw[blue] (-1.7394,1.0118) node[anchor=north east] {{$\gamma_{11}$}};
\filldraw[blue] (-0.2775,0.9999) node[anchor=north east] {{$\gamma_{22}$}};
\filldraw[blue] (2,1) node[anchor=north east] {{$\gamma_{12}$}};
\node at (-1.0571,4.635) {2};
\node at (-4.8038,3.5345) {3};
\node at (2.6945,3.509) {1};
\node at (-4.799,2.9811) {4};
\node at (2.7008,2.9683) {5};
\node at (-4.71,2.0587) {8};
\node at (2.7072,2.0142) {9};
\filldraw[blue]  (-0.1788,4.1878) circle [radius=2pt];
\filldraw[blue]  (-3.9886,3.6277) circle [radius=2pt];
\filldraw[blue]  (-0.7785,2.9957) circle [radius=2pt];
\filldraw[blue]  (-3.9922,2.9957) circle [radius=2pt];
\filldraw[blue]  (-1.256,2.0083) circle [radius=2pt];
\filldraw[blue]  (-3.9922,2.0083) circle [radius=2pt];
\filldraw[blue]  (1.9893,3.5746) circle [radius=2pt];
\filldraw[blue]  (1.9785,2.0054) circle [radius=2pt];
\filldraw[blue]  (1.9822,3.0037) circle [radius=2pt];
\filldraw[blue]  (-2.0496,4.1735) circle [radius=2pt];
\filldraw[blue]  (-0.8575,2.0118) circle [radius=2pt];
\filldraw[blue]  (-1.4068,2.9993) circle [radius=2pt];
\end{tikzpicture}
\quad\quad
\begin{tikzpicture}[baseline={([yshift=-12ex]current bounding box.center)},scale=0.15]
                \draw[black,thick] (0,5)--(-5,5)--(-5,0)--(0,0)--cycle;
                \draw[black,thick] (1.93,5.52)--(0,5)--(0.52,6.93);
                \draw[black,thick] (1.93,-0.52)--(0,0)--(0.52,-1.93);
                \draw[black,thick] (-6.93,5.52)--(-5,5)--(-5.52,6.93);
                \draw[black,thick] (-6.93,-0.52)--(-5,0)--(-5.52,-1.93);
                \filldraw[black] (1.93,6) node[anchor=west] {{$4$}};
                \filldraw[black] (0.52,6.93) node[anchor=south] {{$2$}};
                \filldraw[black] (1.93,-1) node[anchor=west] {{$5$}};
                \filldraw[black] (0.52,-1.93) node[anchor=north] {{$6$}};
                \filldraw[black] (-6.93,6) node[anchor=east] {{$9$}};
                \filldraw[black] (-5.52,6.93) node[anchor=south] {{$1$}};
                \filldraw[black] (-6.93,-1) node[anchor=east] {{$8$}};
                \filldraw[black] (-5.52,-1.93) node[anchor=north] {{$7$}};
                \node at (-2.1034,-8) {$F(2,5,7,9)$};
            \end{tikzpicture}
\caption{Two different one-loop configurations at $n=9$ and their solutions; Intersections on the line $(67)$ are checked to be ordered.}
\label{fig.two-four-mass}
\end{figure}
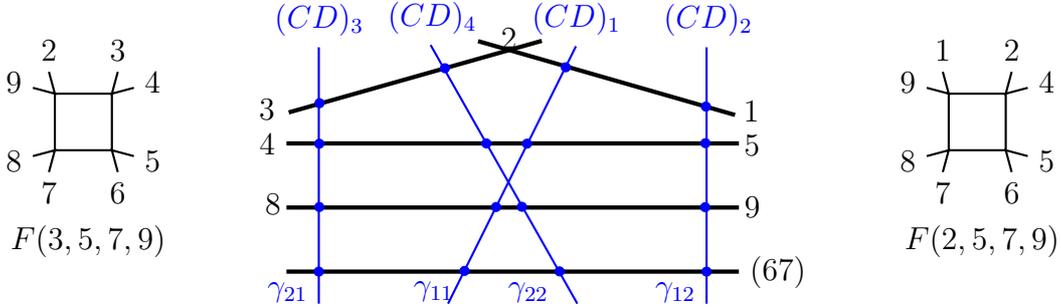

For instance, at $n=9$ we consider Schubert problems from two four-mass boxes $F(2,5,7,9)$ and $F(3,5,7,9)$, whose solutions are called $(CD)_1$ and $(CD)_2$ for the first box and $(CD)_3$ and $(CD)_4$ for the second. Since two boxes share dual points $x_5$, $x_7$ and $x_9$, lines $(45)$, $(67)$ and $(89)$ all intersect with the four lines $\{(CD)_i\}_{i=1\cdots4}$. Now let's consider the four points on the line $(67)$ produced by four $(CD)_i$,   {\it i.e.} considering four points $\gamma_{11}=(67)\cap(89\alpha_{11})$,  $\gamma_{12}=(67)\cap(45\alpha_{12})$, $\gamma_{21}=(67)\cap(89\alpha_{21})$ and $\gamma_{22}=(67)\cap(45\alpha_{22})$ \eqref{gammadelta}, where $\alpha_{1i}$ are the two intersections \eqref{four-mass-solutions} on $(12)$ with $(i,j,k,l)=(2,5,7,9)$, and $\alpha_{2i}$  are the two intersections \eqref{four-mass-solutions} on $(23)$ with $(i,j,k,l)=(3,5,7,9)$. In the region $G_+(4,9)$, we find that the four points are ordered as $\{\gamma_{21},\gamma_{11},\gamma_{22},\gamma_{12}\}$ on $(67)$ (Fig\ref{fig.two-four-mass}), 
and they give a positive letter as
\begin{equation}\label{oddoddletter}
    \mathcal{U}/\mathcal{V}=\frac{(z_{2,5,7,9}-z_{3,5,7,9})(\bar z_{2,5,7,9}-\bar z_{3,5,7,9})}{(z_{2,5,7,9}-\bar z_{3,5,7,9})(\bar z_{2,5,7,9}- z_{3,5,7,9})}
\end{equation}
This letter involves two different square roots $\Delta_{2,5,7,9}$ and $\Delta_{3,5,7,9}$, and has not shown up in known $n=9$ amplitudes yet. However, its appearance was spotted when computing the $9$-point double-box integral \cite{pricom}: 
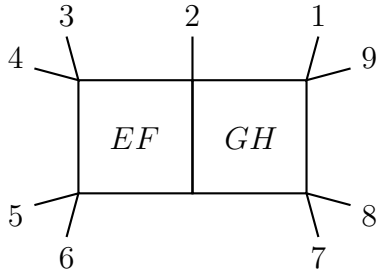
\begin{center}
\begin{tikzpicture}[baseline={([yshift=-.5ex]current bounding box.center)},scale=0.3]
                \draw[black,thick] (0,5)--(-5,5)--(-5,0)--(0,0)--cycle;
                \draw[black,thick] (1.93,5.52)--(0,5)--(0.52,6.93);
                \draw[black,thick] (1.93,-0.52)--(0,0)--(0.52,-1.93);
                \draw[black,thick] (-5,5)--(-5,6.93); 
                \draw[black,thick] (-5,0)--(-5,5)--(-10,5)--(-10,0)--cycle;
                \draw[black,thick] (-11.93,5.52)--(-10,5)--(-10.52,6.93);
                \draw[black,thick] (-11.93,-0.52)--(-10,0)--(-10.52,-1.93);
                \filldraw[black] (1.93,6) node[anchor=west] {{$9$}};
                \filldraw[black] (0.52,6.93) node[anchor=south] {{$1$}};
                \filldraw[black] (1.93,-1) node[anchor=west] {{$8$}};
                \filldraw[black] (0.52,-1.93) node[anchor=north] {{$7$}};
                \filldraw[black] (-11.93,6) node[anchor=east] {{$4$}};
                \filldraw[black] (-10.52,6.93) node[anchor=south] {{$3$}};
                \filldraw[black] (-11.93,-1) node[anchor=east] {{$5$}};
                \filldraw[black] (-10.52,-1.93) node[anchor=north] {{$6$}};
                \filldraw[black] (-5,6.93) node[anchor=south] {{$2$}};
            \node at (-7.543,2.4233) {$EF$};
\node at (-2.4347,2.4234) {$GH$};
\end{tikzpicture}
\end{center}
and similar symbol letters also show up in the $10$-point double-box integral \cite{Kristensson:2021ani}. We believe that such kind of {\it mixed algebraic letters} may also appear in amplitudes at $k+\ell\geq4$ for $n\geq9$. When either one of the two four-mass boxes in \eqref{oddoddletter} degenerates to a lower-mass configuration, mixed algebraic letter degenerates to an original algebraic letter. When both the four-mass boxes degenerate, it comes back to a rational letter.

\paragraph{Mixed algebraic letters with the square root from double-box} By now, we have considered possible cases that combining two different one-loop configurations and considering intersections on external lines. A natural generalization of this idea is that, instead of merely focusing on one-loop configurations, we can take maximal cuts of  arbitrary $\ell$-loop integrals into account. To illustrate this idea, let's take the $9$-point double-box integral as an example again. Explicit computation shows that besides the letter \eqref{oddoddletter}, its alphabet contains another kind of mixed algebraic letters as
\[W_1=\frac{(1{+}a z_{2,5,7,9})(1{+}b \bar{z}_{2,5,7,9})}{(1{+}a \bar{z}_{2,5,7,9})(1{+}b {z}_{2,5,7,9})},\ W_2=\frac{(1{+}a u_1{-}z_{3,5,7,9})(1{+}b u_1{-}\bar{z}_{3,5,7,9})}{(1{+}a u_1{-}\bar{z}_{3,5,7,9})(1{+}b u_1{-}z_{3,5,7,9})}\]
Here we introduce combinations 
\[a=-\frac{1+u_1-u_2+\Delta_9}{2u_1},\ b=-\frac{1+u_1-u_2-\Delta_9}{2u_1}\]
with $\Delta_9=\sqrt{(1-u_1-u_2)^2-4 v_1 v_2}$; and 
\begin{align}
u_1=\frac{\langle1245\rangle\langle6789\rangle}{\langle4589\rangle\langle1267\rangle},\ v_1=\frac{\langle1289\rangle\langle4567\rangle}{\langle4589\rangle\langle1267\rangle},\ 
u_2=\frac{\langle2389\rangle\langle4567\rangle}{\langle4589\rangle\langle2367\rangle},\ 
v_2=\frac{\langle2345\rangle\langle6789\rangle}{\langle4589\rangle\langle2367\rangle}
\end{align}
are four independent cross-ratios of the double-box integral. Although not apparent, we have relation $W_1\leftrightarrow W_2$ under axial symmetry of the integral $(u_1\leftrightarrow u_2,v_1\leftrightarrow v_2)$. Note that $(1{-}u_1{-}u_2)^2{-}4 v_1 v_2$ is again positive definite in the positive region $G_+(4,9)$, as can be checked.

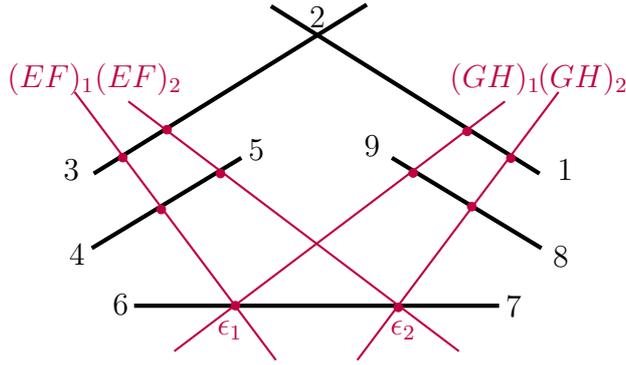
\begin{figure}[htbp]
    \centering
    \begin{tikzpicture}[scale=0.8]
                \draw[black,ultra thick] (-3,1)--(3,1);
                 \draw[black,ultra thick] (-3.6957,1.9535)--(-1.2369,3.4516);
                  \draw[black,ultra thick] (3.6957,1.9535)--(1.2369,3.4516);
                   \draw[black,ultra thick] (3.6638,3.1899)--(-0.75,5.9654);
                   \draw[black,ultra thick] (-3.6638,3.1899)--(0.8239,5.9904);
\draw [purple,thick](-3.9774,4.5424) -- (-0.6694,0.0994);
\draw [purple,thick](-2.3401,0.2458) -- (3.0955,4.3863);
\draw [purple,thick](3.9774,4.5424) -- (0.6694,0.0994);
\draw [purple,thick](2.3401,0.2458) -- (-3.0955,4.3863);
\node at (0.0138,5.7895) {2};
\filldraw[purple]  (-1.3364,1.001) circle [radius=2pt];
\filldraw[purple]  (1.3427,0.9717) circle [radius=2pt];
\filldraw[purple]  (3.1939,3.4457) circle [radius=2pt];
\filldraw[purple]  (2.4755,3.89) circle [radius=2pt];
\filldraw[purple]  (2.5511,2.6422) circle [radius=2pt];
\filldraw[purple]  (1.5869,3.1999) circle [radius=2pt];
\filldraw[purple]  (-2.4591,3.9184) circle [radius=2pt];
\filldraw[purple]  (-3.187,3.4552) circle [radius=2pt];
\filldraw[purple]  (-1.58,3.1905) circle [radius=2pt];
\filldraw[purple]  (-2.5537,2.5949) circle [radius=2pt];
\node at (-4.0323,3.2628) {3};
\node at (4.088,3.2533) {1};
\node at (-3.9381,1.8448) {4};
\node at (-0.9884,3.5842) {5};
\node at (0.9117,3.6503) {9};
\node at (4.0124,1.8164) {8};
\node at (-3.2194,1.0223) {6};
\node at (3.2372,0.994) {7};
\node [purple] at (-4.3727,4.7564) {$(EF)_1$};
\node [purple] at (-2.9329,4.7469) {$(EF)_2$};
\node [purple] at (4.3727,4.7564) {$(GH)_2$};
\node [purple] at (2.9329,4.7469) {$(GH)_1$};
\node [purple] at (-1.4234,0.5995) {$\epsilon_1$};
\node [purple] at (1.4234,0.5995) {$\epsilon_2$};
\end{tikzpicture}
\caption{Two pairs of solutions for $(EF)$ and $(GH)$ from the leading singularity of double-box integral; $(EF)$ intersects with $(23)$, $(45)$, $(67)$ and $(GH)$; $(GH)$ intersects with $(12)$, $(89)$, $(67)$ and $(EF)$. Intersections of $(EF)_i$ with $(GH)_i$ both lie on the line $(67)$, which are call $\epsilon_1$ and $\epsilon_2$ respectively.}
\label{two-loop}
\end{figure}

The square root $\Delta_9$ comes from the leading singularity of $9$-point double-box integral.  Following the procedure in appendix A, there are two pairs of different solutions for loop momenta $(EF)$ and $(GH)$ when we compute its leading singularity, which are four determined lines in momentum twistor space (Fig.\ref{two-loop}). On the support of each pair of solutions, we have $LS\propto \frac1{\Delta_9}$, which is the generation of $\Delta_9$ (see appendix A for more details).

Now we are interested in intersections produced by $(EF)_i$ and $(GH)_i$, especially on $(67)$. A non-trivial fact is that, for each $i=1,2$ the intersection of $(EF)_i$ with $(GH)_i$ precisely lies on $(67)$ (see Fig.\ref{two-loop}), {\it i.e.} for each $i$, three lines $(EF)_i$, $(GH)_i$ and $(67)$ are joint at the same point. we denote these two points as $\epsilon_1=Z_6+e_1 Z_7$ and $\epsilon_2=Z_6+e_2 Z_7$ on $(67)$. Here  $e_1$ and $e_2$ are two solutions of \eqref{A2}$=0$. Now together with four $\gamma_{ij}$ and $6$, $7$, we already have $8$ points on $(67)$. Quite non-trivially, these eight points $\{\gamma_{21},\epsilon_1,\gamma_{11},\gamma_{22},\epsilon_2,\gamma_{12},6,7\}$ are ordered on the line $(67)$, forming an $A_5$ configuration!
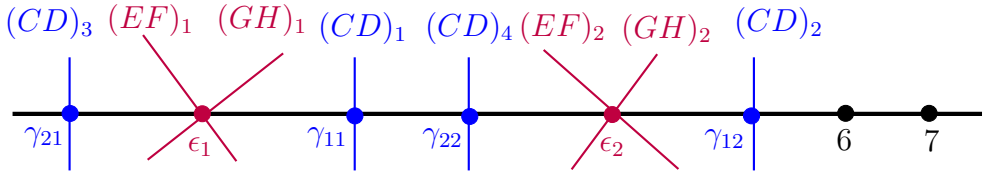
\begin{center}
\begin{tikzpicture}[scale=1.5]
                \draw[black,ultra thick] (-3,1)--(5.5,1);
               
\draw [purple,thick](-1.856,1.697) -- (-1.0372,0.5819);
\draw [purple,thick](-1.8155,0.5945) -- (-0.6275,1.535);
\draw [purple,thick](2.6519,1.5278) -- (1.9017,0.5093);
\draw [purple,thick](2.8382,0.5126) -- (1.6567,1.5652);
\filldraw[purple]  (-1.3364,1.001) circle [radius=2pt];
\filldraw[purple]  (2.2583,0.9941) circle [radius=2pt];
\node [purple] at (-1.7926,1.8267) {$(EF)_1$};
\node [purple] at (1.8193,1.7616) {$(EF)_2$};
\node [purple] at (2.7321,1.7454) {$(GH)_2$};
\node [purple] at (-0.8152,1.8279) {$(GH)_1$};
\node [purple] at (-1.3527,0.709) {$\epsilon_1$};
\node [purple] at (2.2571,0.709) {$\epsilon_2$};
\draw [blue, thick] (0,1.5) -- (0,0.5);
\draw [blue, thick] (1,1.5) -- (1,0.5);
\draw [blue, thick] (-2.5,1.5) -- (-2.5,0.5);
\draw [blue, thick] (3.5,1.5) -- (3.5,0.5);
\filldraw[blue]  (3.479,0.9823) circle [radius=2pt];
\filldraw[blue]  (1.0087,0.977) circle [radius=2pt];
\filldraw[blue]  (0.0015,0.9823) circle [radius=2pt];
\filldraw[blue]  (-2.4952,1.0035) circle [radius=2pt];
\filldraw[black]  (5.036,1.0048) circle [radius=2pt];
\filldraw[black]  (4.3044,1.0048) circle [radius=2pt];
\node [blue] at (0.7721,0.7815) {$\gamma_{22}$};
\node [blue] at (3.2423,0.7762) {$\gamma_{12}$};
\node [blue] at (-2.7213,0.7974) {$\gamma_{21}$};
\node [blue] at (-0.2404,0.7921) {$\gamma_{11}$};
\node [blue] at (-2.6783,1.8045) {$(CD)_3$};
\node [blue] at (0.0692,1.7569) {$(CD)_1$};
\node [blue] at (3.7162,1.8047) {$(CD)_2$};
\node [blue] at (1.0059,1.7621) {$(CD)_4$};
\node at (4.2943,0.7709) {6};
\node at (5.0577,0.7656) {7};
\end{tikzpicture}
\end{center}
Furthermore, two mixed algebraic letters $\{W_1 ,W_2\}$ can be constructed from its $A_1$ sub-configurations $\{\gamma_{21},\epsilon_1,\gamma_{22},\epsilon_2\}$ and $\{\epsilon_1,\gamma_{11},\epsilon_2,\gamma_{12}\}$  and their corresponding $\mathcal{U}/\mathcal{V}$, as can be checked directly.

Most generally, we can consider arbitrary many $\ell$-loop integrals sharing a same dual point $(i{-}1i)$ and compute their leading singularity respectively. After all loop momenta have been located in momentum twistor space\footnote{As revealed in \cite{Bourjaily:2019hmc}, sometimes this procedure may be obstructed if the integral itself is beyond MPL function. Here we restrict our discussion over MPL cases}, they intersect $(i{-}1i)$ at several points. Once these intersections are checked to be ordered (which is quite non-trivial if we can find any such configurations), we can compute cross-ratios of the points. Some of them are expected to give physical singularities of the $\ell$-loop integrals we begin with, like what we have seen from the double-box integral.

\section{Discussions}
In this paper we went through both one-loop and two-loop Schubert problems, which correspond to solving leading singularities of DCI integrals in planar $\mathcal{N}=4$ SYM theory. Solutions of loop momenta became determined lines in momentum twistor space. Besides considering intersections on loop momenta (internal lines), we also considered intersections on external lines when solutions of different Schubert problems intersect with a same line $(i{-}1i)$. We discovered that, when external $\mathbf{Z}$ are evaluated in the positive region $G_+(4,n)$, in each configuration, intersections on a given line were checked to be ordered, and they form an $A_n$ configuration (mainly $A_1$ configurations with four points). This makes cross-ratios of these intersections positive definite. Since these cross-ratios coincide with physical singularities of amplitudes/integrals, we therefore explained the positivity of their symbol letters in the positive region. Especially, from $A_1$ configurations on external lines, we successfully reproduced the $18$ multiplicatively independent algebraic letters for $n=8$ amplitudes. Finally, we also discussed a new kind of mixed algebraic letters at $n\geq9$. As symbol letters of the $9$-point double-box integral, their positivity was also associated to the ordering of intersections on an external line, and we believe they will be symbol letters of planar $\mathcal{N}=4$ SYM amplitudes as well at $k+\ell\geq4$.

Several problems are remained to be solved.  The first and the most important problem is to look for the condition that guarantees intersections to be ordered on a given line. In fact, there exists a two-loop counterexample, where intersections on the solution of Schubert problem are not ordered:
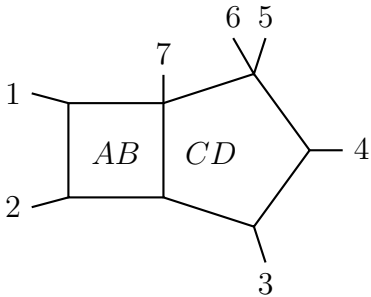
\begin{center}
\begin{tikzpicture}[scale=0.25]
                \draw[black,thick] (0,0)--(0,5)--(4.76,6.55)--(7.69,2.5)--(4.76,-1.55)--cycle;
                \draw[black,thick] (9.43,2.5)--(7.69,2.5);
                \draw[black,thick] (3.67,8.45)--(4.76,6.55)--(5.37,8.45);
                \draw[black,thick] (4.76,-1.55)--(5.37,-3.45);
                \draw[black,thick] (0,5)--(-5,5)--(-5,0)--(0,0);
                \draw[black,thick] (0,5)--(0,6.5);
                \draw[black,thick] (-6.93,5.52)--(-5,5);
                \draw[black,thick] (-6.93,-0.52)--(-5,0);
                \filldraw[black] (5.37,8.45) node[anchor=south] {{$5$}};
                 \filldraw[black] (3.67,8.45) node[anchor=south] {{$6$}};
                \filldraw[black] (0,6.5) node[anchor=south] {{$7$}};
                \filldraw[black] (5.37,-3.45) node[anchor=north] {{$3$}};
                \filldraw[black] (9.43,2.6) node[anchor=west] {{$4$}};
                \filldraw[black] (-6.93,5.52) node[anchor=east] {{$1$}};
                \filldraw[black] (-6.93,-0.52) node[anchor=east] {{$2$}};
                \filldraw[black] (-2.5,3.5) node[anchor=north] {{$AB$}};
                  \filldraw[black] (2.5,3.5) node[anchor=north] {{$CD$}};
            \end{tikzpicture}
\end{center}
Here four intersections on one of the maximal cut solutions of $(CD)$ yields a minor $\langle7(12)(36)(45)\rangle$, which is not a cluster variable, nor positive definite in the positive region!  Moreover, as a two-loop maximal cut, this configuration corresponds to the second $7$-point plabic graph explored in \cite{He:2020uhb}, and associated Yangian letters consist of the same non-cluster variable as well. It is also an interesting problem to systematically reveal the relation between Yangian invariants and Schubert problems.

Secondly, in our construction, rational letters (appear in the amplitudes of arbitrary $k+\ell$), algebraic letters (appear when $k+\ell\geq3$) and mixed algebraic algebraic letters (are supposed to appear when $k+\ell\geq4$) are all from ordered intersections on external lines when combining two one-loop configurations. Although generations of these three kinds of letters are unified in the language of Schubert problems, the reason why mixed algebraic letters are prohibited in  amplitudes at $k+\ell<4$ remains unclear (similarly algebraic letters do not show up when $k+\ell<3$).  Moreover, whether there are more complicated algebraic letters in the alphabet of amplitudes at higher $k+\ell$? Can we uncover them from certain Schubert problems?

Finally, in this note we only focus on leading singularities of MPL integrals. It is also possible for us to extend the discussion to elliptic cases, for instance, the $10$-point double-box integral \cite{Kristensson:2021ani}. The difference is that now loop momenta are no longer determined lines in momentum twistor space, since to solve the elliptic leading singularity, the last unfixed freedom of loop momenta is taken an contour integration \cite{Bourjaily:2020hjv}, instead of being determined by a specific residue. It would be extremely interesting if cross-ratios from this ``elliptic Schubert problem" still offer us any physical information about the $10$-point double-box integral, and we leave it for future study.

\appendix
\section{Schubert problems and leading singularities}
In this appendix we present some details in finding solutions of Schubert problems from computing corresponding leading singularities of integrals. Recall that to obtain the leading singularity of an $\ell$-loop integral, we need to solve $4\ell$ on-shell conditions from the integrand  and take corresponding $4\ell$-fold residues for loop momenta. We will see that through this approach, geometrically we in fact locate loop momenta as determined lines in momentum twistor space. Therefore the corresponding Schubert problem is solved.

Back to the one-loop four-mass box configuration $F(i,j,k,l)$ again. We are looking for the solutions for loop momentum $(AB)_i$ satisfying $4$ on-shell conditions $\langle ABi{-}1i\rangle=\langle ABj{-}1j\rangle=\langle ABk{-}1k\rangle=\langle ABl{-}1l\rangle=0$. As a geometric problem in $\mathbb{P}^3$, projectively we can always parametrize two points $A$ and $B$ by four independent momentum twistors, which following the choice in \cite{Arkani-Hamed:2010pyv} are decided to be
\[A=\alpha Z_i+\epsilon Z_{j{-}1}+Z_{i{-}1}, B=\rho Z_i+ \beta Z_{j{-}1}+ Z_j\]
after a $GL(2)$ gauge-fixing. Therefore conditions $\langle ABi{-}1i\rangle=\langle ABj{-}1j\rangle=0$ indicate $\epsilon=\rho=0$, while $\langle ABk{-}1k\rangle=\langle ABl{-}1l\rangle=0$ result in quadratic equations of $\alpha$ and $\beta$ as
\[\alpha=-\frac{\beta\langle i{-}1j{-}1k{-}1k\rangle+\langle i{-}1jk{-}1k\rangle}{\beta\langle ij{-}1k{-}1k\rangle+\langle ijk{-}1k\rangle},\ \  \beta=-\frac{\alpha\langle i{-}1jl{-}1l\rangle+\langle ijl{-}1l\rangle}{\alpha\langle i{-}1j{-}1l{-}1l\rangle+\langle ij{-}1l{-}1l\rangle}\]
which yield two solutions $\alpha_1$ and $\alpha_2$ for $A$ (and $\beta_1$ and $\beta_2$ for $B$ correspondingly) as \cite{Bourjaily:2013mma}:
\begin{equation}\label{four-mass-solutions}
\begin{split}
\alpha_1&=Z_i{+}Z_{i{-}1}\frac{\langle ij^\prime j(kk^\prime\cap(l^\prime li^\prime))\rangle+\langle i^\prime j^\prime j(kk^\prime\cap(l^\prime li)\rangle+\langle ii^\prime kk^\prime\rangle\langle j^\prime jl^\prime l\rangle\Delta_{i,j,k,l}}{2\langle j^\prime j(kk^\prime \cap(l^\prime li^\prime )i^\prime\rangle}\\
\alpha_2&=Z_{i{-}1}+Z_i\frac{\langle i^\prime ll^\prime (k^\prime k)\cap(jj^\prime i)\rangle+\langle ill^\prime(k^\prime k)\cap(jj^\prime i^\prime)\rangle+\langle i^\prime i k^\prime k\rangle\langle jj^\prime ll^\prime\rangle\Delta_{i,j,k,l}}{2\langle ll^\prime(k^\prime k)\cap(jj^\prime i)i\rangle}\\
\beta_1&=Z_{j{-}1}+Z_{j}\frac{\langle j^\prime ii^\prime(l^\prime l)\cap(kk^\prime j )\rangle+\langle ji i^\prime(l^\prime l)\cap(kk^\prime j^\prime)\rangle+\langle ii^\prime kk^\prime\rangle\langle j^\prime jl^\prime l\rangle\Delta_{i,j,k,l}}{2\langle ii^\prime (l^\prime l)\cap(kk^\prime j )j\rangle}\\
\beta_2&=Z_j+Z_{j{-}1}\frac{\langle j k^\prime k (l l^\prime)\cap(i^\prime i j)\rangle+\langle j^\prime k^\prime k(l l^\prime)\cap(i^\prime i j^\prime)\rangle+\langle i^\prime i k^\prime k\rangle\langle jj^\prime ll^\prime\rangle\Delta_{i,j,k,l}}{2\langle k^\prime(l l^\prime)\cap(i^\prime i j)j\rangle}
\end{split}
\end{equation}
where we abbreviate $a^\prime:=a{-}1$, and $\Delta_{i,j,k,l}$ reads \eqref{delta}. Intersections $\gamma$ on $(k{-}1k)$ and $\delta$ on $(l{-}1l)$ can then be worked out from the geometric viewpoint as
\begin{equation}\label{gammadelta}\gamma_i=(k{-}1k)\cap(l{-}1l\alpha_i),\ \  \delta_i=(l{-}1l)\cap(k{-}1k\beta_i)
\end{equation}
Finally, taking residue of the integrand \eqref{integrand1} at the solution of loop momentum, we will see $LS\propto\frac1{\Delta_{i,j,k,l}}$ as expected.

Let's move to the more non-trivial two-loop example, {\it i.e.} $9$-point double-box integral and its leading singularity. In the geometrical point of view, solving on-shell conditions is in equivalence to locating two lines $(EF)$ and $(GH)$ in $\mathbb{P}^3$, where $(EF)$ intersects with $(GH)$, $(23)$, $(45)$ and $(67)$, and $(GH)$ intersects with $(EF)$, $(12)$, $(67)$ and $(89)$. Similar to the one-loop case, we parametrize four momentum twistors $\{E,F,G,H\}$ as
\begin{align}\label{A1}
E&=\alpha_1 Z_5+\beta_1 Z_6+Z_4,\ F=\alpha_2 Z_5+\beta_2 Z_6+Z_7\nonumber\\
G&=\gamma_1 Z_7+ \delta_1 Z_8+Z_9,\ H=\gamma_2 Z_7+ \delta_2 Z_8+Z_6 
\end{align}
and determine $8$ variables by on-shell conditions from the integral. However, the integral only has $7$ propagators. Therefore equations from on-shell conditions are not enough for us to determine all the variables. In fact, after putting all $7$ propagators on shell, we have relations 
\begin{align}\label{A11}
    \alpha_2&=0,\ \beta_1=0,\ \gamma_1=0,\ \delta_2=0 \nonumber \\
    \delta_1&=-\frac{\langle1269\rangle{+}\gamma_2\langle1279\rangle}{\langle1268\rangle{+}\gamma_2\langle1278\rangle},\ \alpha_1=-\frac{\langle2346\rangle+\gamma_2\langle2347\rangle}{\langle2356\rangle+\gamma_2\langle2357\rangle},\ \beta_2=\frac1{\gamma_2}
\end{align}
between variables, and are left with a quadratic factor
\begin{equation}\label{A2}
\begin{split}
\langle7\ (89)\cap(612)\ (645)\cap(623)\rangle-\langle6\ (89)\cap(712)\ (745)\cap(723)\rangle\gamma_2^2\\
+(\langle6\ (45)\cap(236)\ (789)\cap(712)\rangle-\langle7\ (45)\cap(237)\ (689)\cap(612)\rangle)\gamma_2
\end{split}
\end{equation}
on the denominator. Relation $\beta_2=\frac1{\gamma_2}$ together with $\alpha_2=\delta_2=0$ guarantees that $F=\frac1{\gamma_2}H$. Projectively they are identical on $(67)$.  Therefore intersections of $(EF)$ with $(GH)$ always lie on $(67)$. 

In principle, once satisfying \eqref{A11},  for arbitrary $\gamma_2$ in these expressions, the corresponding two lines $(EF)$ and $(GH)$ from \eqref{A1} are solutions of this two-loop Schubert problem. We are mainly interested in the solutions that correspond to leading singularity of this integral; we need to set $\gamma_2$ at the zero points of \eqref{A2}. Note that discriminant $\Delta_9^\prime$ of quadratic equation \eqref{A2}=0 is related to the square root $\Delta_9$ as
\[\Delta_9^\prime=\langle2367\rangle\langle4589\rangle\langle1267\rangle\Delta_9,\]
and the resulting two pairs of solutions $(EF)_i$ and $(GH)_i$ for $i=1,2$ are the solutions we used in section $4$ to construct letters $W_1$ and $W_2$. Finally, taking residues at the solutions to derive the leading singularity, we see $LS\propto\frac1{\Delta_9}$.

\section*{Acknowledgement}
The author would like to thank Nima Arkani-Hamed for proposing the problem and for inspiring discussions. We thank Song He for guidance during the project, and Zhenjie Li, Yichao Tang for helpful discussions and comments on this manuscript. We also thank Chi Zhang for sharing explicit result of the integral and for collaborations on related projects. This research is supported in part by National Natural Science Foundation of China under Grant No. 11935013,11947301,12047502,12047503.

\bibliographystyle{utphys}
\bibliography{main}
\end{document}